\documentclass[reprint,prx,amsmath,amssymb,floatfix]{revtex4-2}

\usepackage{graphicx}% Include figure files
\usepackage{dcolumn}% Align table columns on decimal point
\usepackage{bm}% bold math
\usepackage{float}
\usepackage{import} % for including the supp in the main
\usepackage{natbib}
\usepackage{xcolor}
\usepackage[normalem]{ulem}
\usepackage{xr}

\newcommand{\prx}[1]{\textcolor{black}{#1}}
\newcommand{\sprx}[1]{}

\begin{document}

\title{Electrically controlled photonic circuits of field-induced dipolaritons with huge nonlinearities}
%  
%*************************%
%% 
\author{Dror Liran}
\email{dror.liran@mail.huji.ac.il}
\author{Ronen Rapaport}
\affiliation{Racah Institute of Physics, The Hebrew University of Jerusalem,  Jerusalem 9190401, Israel}
\author{Jiaqi Hu}
\author{Nathanial Lydick}
\author{Hui Deng}
\affiliation{University of Michigan, Ann Arbor, MI 48109, USA}
\author{Loren Pfeiffer}
\affiliation{Department of Electrical Engineering, Princeton University, Princeton, NJ, 08544 USA}
%*************************%

\date{\today}

\begin{abstract}
Electrically controlled photonic circuits hold promise for information technologies with greatly improved energy efficiency and quantum information processing capabilities. However, weak nonlinearity and electrical response of typical photonic materials have been two critical challenges. Therefore hybrid electronic-photonic systems, such as semiconductor exciton-polaritons, have been intensely investigated for their potential to allow higher nonlinearity and electrical control, with limited success so far. Here we demonstrate an electrically-gated waveguide architecture for dipolar-polaritons that allows enhanced and electrically-controllable polariton nonlinearities, enabling an electrically-tuned reflecting switch (mirror) and transistor of the dipolar-polaritons. The polariton transistor displays blockade and anti-blockade by compressing a dilute dipolar-polariton pulse exhibiting very strong dipolar interactions. The large nonlinearities are explained using a simple density-dependent polarization field that very effectively screens the external electric field, with an order of magnitude enhancement of the nonlinearities as compared to fixed dipoles. We project that a quantum blockade at the single polariton level is feasible in such a device.
\end{abstract}

\maketitle

\section{Introduction}
Photons are excellent carriers of information. Bits or qu-bits can be encoded on their different degrees of freedom, guided in complex on-chip integrated light circuits, and travel macroscopic distances with minimal loss or decoherence \cite{Wang2019IntegratedTechnologies,Elshaari2020HybridCircuits,Bogaerts2020ProgrammableCircuits}. Yet photons do not interact with each other nor with external electric and magnetic fields. Therefore it has been a challenge to use them for information processing, such as gates and logic operations, or to integrate them with electronic control and other electronic-based states.
Therefore, integrated logic circuits are still mainly made of purely electronic states, either free-carriers or new attempts for circuits based on bare excitons \cite{butovReview2017,ExcitonRevTMD2022,SAWExcitons2014,ButovTransistor2008}. These approaches are limited in their operation speed and in their ability to transfer information coherently.
A promising route to induce photon-photon interactions is to strongly couple a confined photon with a resonant electronic excitation to create a light-matter superposition state known as a polariton \cite{Deng2010,SanvittoTheRoadDevices,Chang2018Colloquium:Photons}.  Polaritons can preserve the photonic information while enabling effective photon-photon interactions and interactions with external fields via their matter constituent.
In atomic platforms, laser-controlled quantum operations utilizing Rydberg polaritons have been successfully demonstrated \cite{Peyronel2012QuantumAtoms,Tiarks2018AInteractions}, but on-chip integration or electrical control are very challenging.
In semiconductor platforms, microcavity exciton-polaritons have been used to demonstrate optically controlled switches \cite{De2023Room-TemperatureCondensates,Zasedatelev2021Single-photonTemperature,SpinSwitchAmo2010}, routers \cite{Marsault2015,TunnelingDiode2013Bloch} and transistors \cite{Gao2012Transistor,Ballarini2013,Suchomel2017PrototypeSwitch}, as well as a few percent of anti-bunching\cite{Delteil2019,Munoz-Matutano2019EmergencePolaritons}, albeit only through inefficient optical control and only in relatively bulky vertical or free-space microcavities that are still challenging for integration.
Moreover, the nonlinearity of these polariton systems, even though much stronger than conventional optical materials remain limited due to the contact-like interactions of polaritons \cite{Directinteractions_snoke_2017,Snoke2023ReanalysisMicrocavities,Vladimirova2010interactions}. 
Alternatively, exciton-polaritons in waveguide structures have recently been shown to feature nonlinearities over an order-of-magnitude greater than their microcavity counterpart \cite{Rosenberg2016,Rosenberg2018a,Suarez-Forero2021EnhancementInteractions}. 
In this system, electrical gates can be placed in the proximity of the waveguide to polarize the excitonic component of the polaritons, creating dipolar-polaritons, or dipolaritons, which interact with each other strongly through the long-range dipole-dipole interactions \cite{Rosenberg2016,Rosenberg2018a,Suarez-Forero2021EnhancementInteractions,Tsintzos2018ElectricalCondensates,Cristofolini2012,Togan2018EnhancedPolaritons,Datta2022HighlyMoS2}.
The waveguide dipolaritons thus hold great promise for constructing electrically-controlled complex quantum-light circuitry on-chip \cite{Walker2017,Liran2018a,Brimont2020GaN,SuarezLaser2020,Nigro2022IntegratedInterferometry,Christensen2022MicroscopicDipolaritons,Ling2023DeeplyNanophotonics}, yet no such demonstration has been made.

In this work, we make an important step towards polariton-based quantum circuitry, by using sectioned electrically-gated waveguide devices, and demonstrate a reflecting polariton electrical switch and a polariton transistor, operating with a very low number of photons.
We first show that fast propagating polaritons in a sectioned gated waveguide device can be very efficiently blocked and reflected
by an electrically-induced mismatch in the polariton density of states, thus realizing an electrical Stark switch for light. We then show a device that acts as an electrically-tuned polariton optical transistor, displaying both a blockade and an anti-blockade for a dilute polariton pulse.
The switch and transistor display an exceptional extinction ratio up to $>20$dB. The apparent large non-linearity of the very dilute polariton pulse is due to a combination of the strong field-induced repulsive dipolar interactions \cite{Rosenberg2018a}, together with a drastic slow-down of the polariton speed under the electrical gate, where the polaritons suddenly become dipolar and exciton-like and are "squeezed" into a high-density, strongly interacting pulse. The strong nonlinearities are  explained by an effective field-induced dipolar screening model, elucidating the large interaction enhancement of induced dipolaritons compared to non-polar polaritons and fixed-length dipolaritons. We project that a true two-polariton blockade is within close reach in such a system.

\section{Electrical switch for waveguide dipolaritons}
%*************************%
%%          Figure 1     %%
\begin{figure*}[ht]
    \centering
    \includegraphics[width = \textwidth]{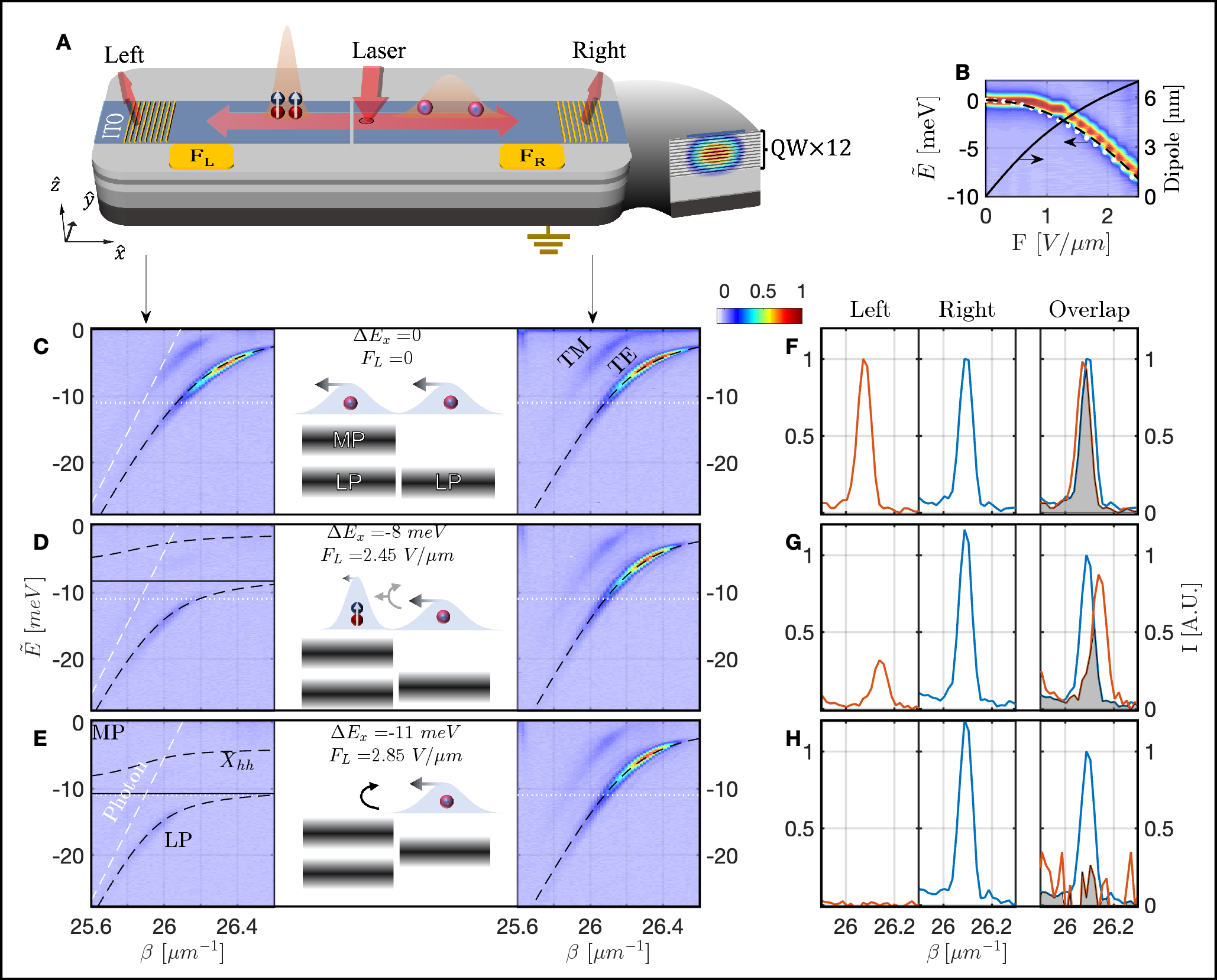}
    \caption{
    \textbf{Electrically Polarized Polaritons in a split-gate waveguide} % title
    \newline
    (\textbf{A}) Illustration of the device: a slab waveguide with 12 QWs in its core, the ITO strip defines the optical mode laterally and serves as a top electrode. The channel is split, allowing an independent electrical field in each section ($F_R / F_L$). The polaritons are injected into the center of the channel, propagate to both directions and couple out through the gratings. 
    (\textbf{B}) The energy of $X_{hh}$ as a function of $F$.
    The white dots are the numerically calculated exciton energies (see SM), and the dashed black line is a quadratic fit $\Delta E_X = -\alpha F^2$ with $\alpha =1.3 meV/(V^2/\mu m^2)$.
    The black solid line plots the calculated electric dipole length, (right axis).
    (\textbf{C-E}) Measured dispersion spectra from the two sides of the channel, for three different values of $\Delta E_{X}(F_{L},F_{R}=0)$. The LP/MP fits are plotted with dashed black lines, whereas $X_{hh}$ and the photon are marked by solid black and dashed white lines, respectively. The middle column presents schematics of the step-potential effect, the increase in $\Delta E_{X}$ decreases the overlap of the LP in the two sections, decreasing the transmitted signal, until near blockage.
   \textbf{F-H} Cross-sections of the spectra, at  $\tilde{E}(|\beta|) = -11\, meV$, marked in (C-E) by dotted white lines. The left (right) side corresponds to the transmission (reflection). The right-most column shows overlaid, normalized cross-section pairs with gray-filled overlaps.
    } % caption
    \label{fig:Fig1}
\end{figure*}
%*************************%
Our first device, a split-gate waveguide, illustrated in Fig. \ref{fig:Fig1}A, is designed to realize an electrostatic potential-step for the polaritons and to demonstrate an electrically-controlled Stark switch for polaritons.  The device is fabricated by depositing a 20-micron wide and 200-micron long Indium-Tin-Oxide (ITO) strip on top of an AlGaAs slab waveguide (WG) sample containing multiple GaAs quantum wells in its core. The full details of the sample, which is similar to the one in our previous works \cite{Rosenberg2016,Rosenberg2018a,Liran2018a}, can be found in the SM. The strong interaction between the heavy-hole exciton $X_{hh}$, the light-hole exciton $X_{lh}$, and the Transverse-electric (TE)  WG-photon results in 3 polariton modes: lower polariton (LP), middle polariton (MP) and an upper polariton \cite{Rosenberg2016} (the polariton modes resulting from the Transverse magnetic (TM) WG-mode, seen in Fig. \ref{fig:Fig1}C are not discussed in this work). The dispersion relation of the LP/MP along the bare modes ($X_{hh}$/photon) are plotted on top of measured spectra in Fig. \ref{fig:Fig1}C-E. The ITO strip laterally confines the optical modes \cite{Liran2018a}, and also serves as the top electrode with respect to the doped substrate. In this design, the top ITO electrode is split, and has a $1 \, \mu m$ gap in the middle, and the two resulting sections are independently biased with respect to the common bottom electrode. 
The vertically aligned electric field ($F$, see Eq. \ref{eq:supp:V2F} in Appendix \ref{app:Sceening model}) under each section polarizes the exciton constituent of the polariton and results in dipolaritons with electric dipole moments along the z-axis \cite{Rosenberg2016,Rosenberg2018a}. The induced dipole-moment then interacts with the same electric field resulting in a Stark-shift of the exciton, $\Delta E_X(F)=X_{hh}(F)-X_{hh}(0) = -\alpha F^2$, where alpha is the polarizability, as is seen in Fig. \ref{fig:Fig1}B. This stark-shift of the exciton and thus of the whole polariton dispersion as well as the polariton dipole moment can be controlled independently in each section by the corresponding applied bias. The polariton energies here after are quoted relative to $X_{hh}(0) = 1527 meV$ as: $\tilde{E}(|\beta|)=E(|\beta|)-X_{hh}(0)$, where $\beta$ is the WG-polariton propagation wavevector.

A schematic description of an experiment, done at $T\simeq 5K$, is shown in Fig. \ref{fig:Fig1}A. The WG-polaritons are injected to right channel $\sim 15$ microns from the ITO gap, using a non-resonant pulsed laser ($\lambda_p=774\, nm$, pulse duration $\tau_p=94 \, ps$, and a repetition rate $200 \, kHz$) and then propagate towards the two ends of the channel and couple out through left and right grating couplers.

Figure \ref{fig:Fig1}C presents the dispersion of the polaritons emitted from the left and right grating couplers, taken at a flat potential, where $F_L=F_R=0$. The spectrum shows a similar dispersion and occupation of LPs from the two sides of the channel, as expected from symmetry. It also shows that the gap in the ITO has a negligible effect, and that left-moving LPs do not experience a significant reflection or loss compared to right movers, as they experience a continuous potential landscape and an identical dispersion on both sections, $\tilde{E}_R(|\beta|)=\tilde{E}_L(|\beta|)$.

\begin{figure}[!ht]
    \centering
    \includegraphics[width = 8.6 cm]{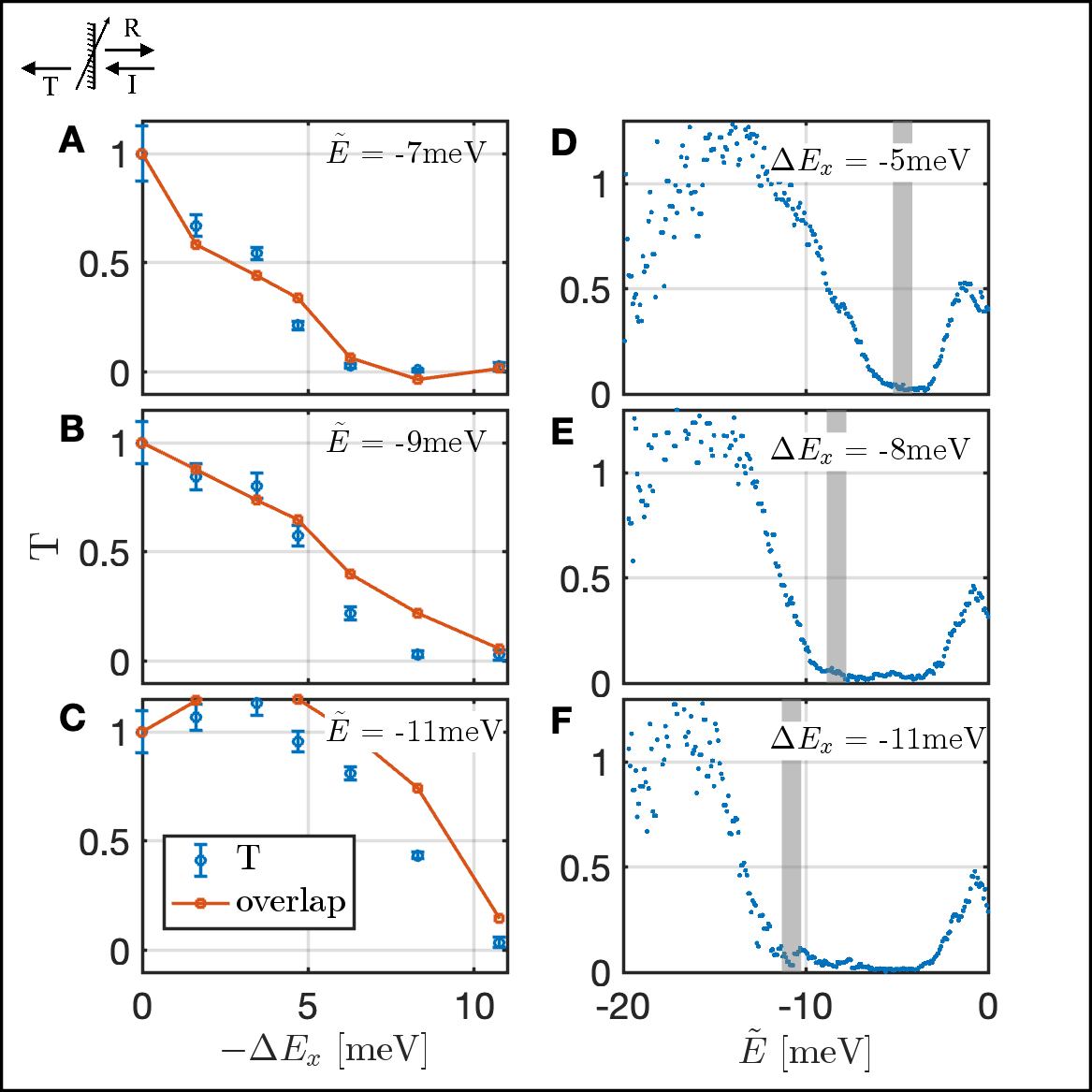}
    \caption{
    \textbf{A tunable polariton electrical mirror with a voltage-controlled potential step}
    \newline
    (\textbf{A-C}) Transmission of the left moving polaritons as a function of $-\Delta E_X(F_L)$, for different energy cuts ($\tilde{E}=-7,-9,-11 \, meV$ respectively), the transmission (blue symbols) is compared against the overlap of the states in the two sides of the potential discontinuity (red dots). The error-bars are the standard deviation.
    (\textbf{D-F}) Transmission of the left moving polaritons for different step size ($\Delta E_X(F_L) = -5,-8,-11 \, meV$ respectively). The energy of the exciton is marked by a vertical gray line.
    The signal at energies near $\tilde{E} = 0$ results from the residual bare exciton emission at the excitation point leaking into the field of view of the collection optics.
    } % caption
    \label{fig:Fig2}
\end{figure}

The situation is modified when a finite $F_L$ is applied to the left section: the left LP and MP modes become red-detuned with respect to the right LP and MP modes, creating a dispersion mismatch for the left movers $\tilde{E}_R(|\beta|)\neq\tilde{E}_L(|\beta|)$. This has an increasingly dramatic effect on the dynamics as seen in Fig. \ref{fig:Fig2}D,E: while the right movers seem to have a slight increase in population, the transmission of the left movers from the left output coupler becomes much weaker for all energies, with essentially zero transmission for all states above the corresponding bare exciton energy.

This effect is explained qualitatively in the schematics in the middle column of Fig. \ref{fig:Fig1}C-E: While the polariton dispersion at the right section is unchanged, in the left section it becomes increasingly red-shifted with increased $F_L$, and the polaritons become dipolar \cite{Rosenberg2018a}. As a result, the $\beta$ mismatch of the LP modes with a given energy in the right and left sections increases, creating an increased discontinuity in $\beta(\tilde{E})$ for the left movers at the intersection between the two sections. This also creates a discontinuity in the polariton group velocity, $v_g$, where polaritons moving to the left section experience a sudden decreased $v_g$, i.e., a slow down, in particular for polaritons with $\tilde{E}_R(|\beta|)\simeq\tilde{E}_X(F_L)$. This discontinuity results in a decreased LP transmission and an increase in reflection. Strikingly, left movers with energies $\tilde{E}_R(|\beta|)$ above  $\tilde{E}_X(F_L)$ have no polariton states in the left section, as they lie in the Rabi energy gap between the LP and the MP branches (LP-MP gap), so they experience zero transmission as clearly seen in the left spectra in Fig.\ref{fig:Fig1}D,E.

In Fig. \ref{fig:Fig1}F-H we plot a constant energy cross-section of normalized emitted intensity from the left and right couplers $I_{R,L}(\tilde{E} = -11 \, meV,|\beta|)$. The decrease in the transmission and increase in reflection with increasing $F_L$ can be clearly seen, as well as the mismatch in the propagation constants $\beta$ on both sides. The rightmost column plots the signal from the two sides overlaid, each normalized by its own integral.  The relative shift in $\beta$ and the decrease in the overlap of the two sides (gray filling) is clearly seen.
When $\tilde{E}_R(|\beta|)$ lies in the LP-MP gap there is essentially zero overlap and zero transmission, as seen in Fig. \ref{fig:Fig1}H.
%*************************%
%%          Figure 2     %%
%*************************%

% \onecolumngrid
%*************************%
%%          Figure 3     %%
\begin{figure*}
    \centering
    \includegraphics[width = 17cm]{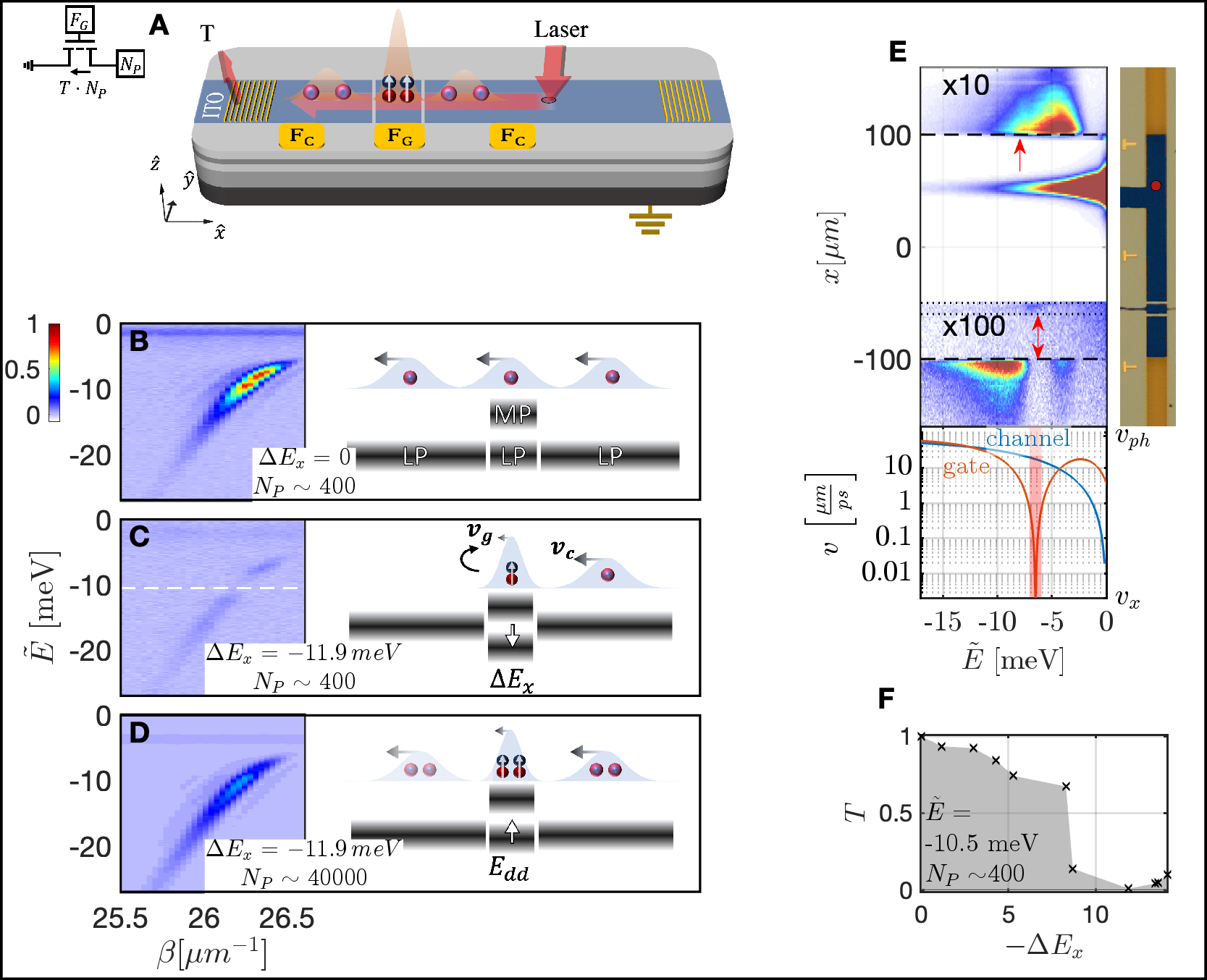}
    \caption{
    \textbf{A dipolariton transistor: strong dipolar interactions of ultra-slow polaritons}
    \newline
    (\textbf{A}) Illustration of the second device.
    The principle of operation is described in \textbf{B,C, and D}:
        (\textbf{B}) Flat potential with a relatively dilute polariton pulse fully transmits;
        (\textbf{C}) Biased gate causes the LP in the channel to overlap mostly with the LP-MP gap in the gate, resulting in a blockage;
        (\textbf{D}) $N_P$ is increased. This results in a compressed pulse of slow polaritons under the gate. The repulsive dipolar interactions ($E_{dd}$) induce a blue-shift of the dispersion under the gate, enabling transmission.
    (\textbf{E})
        (right) Microscope image of the device: the two Au gratings are at the top and the bottom, and the red dot indicates the excitation spot.
        (left) Measured spatially resolved spectrum: Top emission ($x> 100 \mu m$) of the right movers with a flat potential, bottom emission ($x< -100 \mu m$) from left movers propagating through a biased gate ($\Delta E_x  = -7.5$). The bottom spectrum shows a transmission gap - marked by a double red arrow, which also points to a corresponding emission under the gate at the same energy.
        The Middle ($x\simeq 50 \mu m$) section shows the emission of the uncoupled exciton under the excitation spot.
        Attached below is the calculated group velocity ($v_g(\tilde{E})$), of the LP in the channel (blue) and the LP/MP in the gate (red). The red area marks the energy region where $v^G_g\ll v^C_g$, and the gate polaritons become ultra-slow.
    (\textbf{F}) $T$ as a function of $\Delta E_X(F_G)$. The sharp ON-OFF transition of $T$ happens with $\Delta F_G<0.5 V/\mu m$.
    }
    \label{fig:Fig34}
\end{figure*}
%*************************%
% \twocolumngrid

Fig. \ref{fig:Fig2}A-C plots the decreasing transmission ($T$) of 3 different LP energies (integrated over all $\beta$), with increasing Stark shift, $-\Delta E_X(F_L)$. Remarkably, it agrees well with the measured overlap between the polariton states in both sections, as was plotted in Fig. \ref{fig:Fig1}F-H. This validates that the transmission process through the discontinuity is coherent, conserving energy and momentum.  

In Fig. \ref{fig:Fig2}D-F we present $T(\tilde{E}_R)$ for 3 different values of $F_L$. An electrically tunable zero transmission gap with a very high extinction ratio is observed, demonstrating that the polariton optical transmission can be selectively tuned in a continuous manner by low-voltage electrical gating, thus facilitating a tunable electro-optical Stark switch for light.

\section{Electrically controlled optical transistor for dipolaritons}

Next, based on the concepts of our first device, we turn to our second device, illustrated in Fig. \ref{fig:Fig34}A, designed to realize an electrically controlled polariton transistor, and to demonstrate a blockade and an anti-blockade for a dilute polariton pulse, utilizing enhanced dipolar interactions of ultra-slow polaritons. The device, 200-micron long, is divided into three sections: a short 10-micron section positioned 50-micron from the left grating - "the gate", and its two sides - "the channel". The two channel sections are held at $F_C=0$ at all times, while a field $F_G$ under the gate creates a local stark-shift of $X_{hh,G}$: ($\Delta E_X = X_{hh}(F_G) - X_{hh}(0)$). Low-density polariton pulses, each containing $N_P\simeq 400$ polaritons (see SM for details) are non-resonantly injected through the right channel section, either $50$ or $100$ microns away from the gate section, and then propagate in the two directions. Here we only consider left movers that pass through the gate section.
\\
Figure \ref{fig:Fig34}B presents the dispersion of the polaritons emitted from the left grating for a flat potential $F_G=0$. This is used as a reference. 
When the gate is biased, the LP and MP states in the gate are red-shifted with respect to those of the channel, as presented in the schematics of Fig. \ref{fig:Fig34}C. This electrically-induced discontinuity blocks the dilute pulse of left movers with energies at the middle of the LP-MP gap, $\tilde{E} = -10.5 meV$ (see white dashed line), with a maximal extinction ratio $>20dB$, limited only by the measurement SNR.
Fig. \ref{fig:Fig34}F plots the transmission for polaritons with $\tilde{E} = -10.5 \, meV$, as a function of $-\Delta E_X$, again displaying an electrical Stark-switch behavior for the polaritons, but with an even sharper on-off switching field, $\Delta F_G \simeq 0.5 V/\mu m$, due to the well-type double discontinuity in the effective potential, compared to the step-shaped single discontinuity in the first device. Interestingly, the transmission starts to increase again with a further increase of $F_G$ when the LP in the channel coincide with the MP in the gate. This effect will become of importance later for demonstrating a blockade of dipolaritons. 
\\
Fig. \ref{fig:Fig34}E shows the spatially resolved spectrum from the gated device (with $\Delta E_X(F_G) =  - 7.5 \, meV$). A blocking in the transmission spectrum is also clearly observed at the bottom grating at energies corresponding to the LP-MP energy gap under the biased gate (red double arrow), similar to the first device. This blocking of the transmission through the gate is accompanied by an increase in emission from the right coupler at the corresponding energies (red single arrow), indicating that left movers were at least partially reflected from the gate and became right movers. 
The bottom of Fig. \ref{fig:Fig34}E shows the group velocities $v_g^{G,C}(\tilde{E})$ of the polaritons under the gate and the channel respectively, the group velocity is calculated from the derivative of the polariton energy dispersion with respect to the momentum (Eq. \ref{eq:sup:group_vel}). While $v_g^G\simeq v_g^C\simeq v_{ph}$ at low energies, where the LP are photon-like, there is a growing mismatch at higher energies. Remarkably, at polariton energies approaching the LP-MP gap in the gate, $v_g^G$ dramatically drops towards the bare $X_{hh}$ velocity $v_X$, resulting in an extreme slow-down of polaritons under the gate, as large as $v_{X}/v_{ph}\sim 10^{-4}$. 
Interestingly, a weak emission from under the gate (at $x\sim -50 \mu m$) is seen (marked by the double red arrow) at exactly the energy range of these ultra-slow dipolaritons. We thus attribute this emission to ultra-slow dipolaritons which are effectively scattered out of the WG modes through mutual dipolar interactions \cite{Rosenberg2018a}.
\\
Next, we measure the effect of increasing $N_P$ on the transmission. An example is shown in Fig. \ref{fig:Fig34}D for $N_P\simeq 40000$, where the transmission signal recovers almost fully, even at the energies where it was essentially zero at low $N_P$. 
This strongly non-linear transistor-like behavior is explained in the schematics: As $N_P$ increases, the repulsive dipolar interactions of the electrically polarized polaritons penetrating under the gate results in a blue shift $\Delta E_{dd}=g_{dd} \cdot n_G$ of the dipolaritons, where $g_{dd}(F_G)$ is the dipolar interaction constant \cite{Rosenberg2018a} and $n_G$ is the polariton density under the gate.
This interaction induced blue-shift screens the red-shift discontinuity $\Delta E_X$, until $\tilde{E}_G(\beta)\simeq \tilde{E}_C(\beta)$. As a result, the electrical blocking is removed, as the interacting dipolaritons overcome the potential well.
Importantly, the interaction-induced screening effects with increasing $N_P$ are strongly enhanced not only by their dipolar nature \cite{Rosenberg2016,Rosenberg2018a,Suarez-Forero2021EnhancementInteractions,Datta2022HighlyMoS2,Togan2018EnhancedPolaritons,Snoke2021, Christensen2022MicroscopicDipolaritons}, but also by the dramatic slow-down of the polaritons under the gate, $v^G_g\ll v^C_g$, resulting in a spatial compression of the polariton pulse, which for a fixed $N_P$ increases $n_G$ compared to $n_C$
\begin{equation}
        n_G = \int\textbf{d}E\frac{N_P(E)}{\tau_P\cdot w \cdot v_g^G(E)}
    \label{eq:gate_dens}
\end{equation}
where $w$ is the effective width of the optical mode in the lateral direction ($\hat{y}$) (more details on this can be found in the SM.)
% \onecolumngrid
%*************************%
%%          Figure 4    %%
\begin{figure*}
    \centering
    \includegraphics[width = 17.9cm]{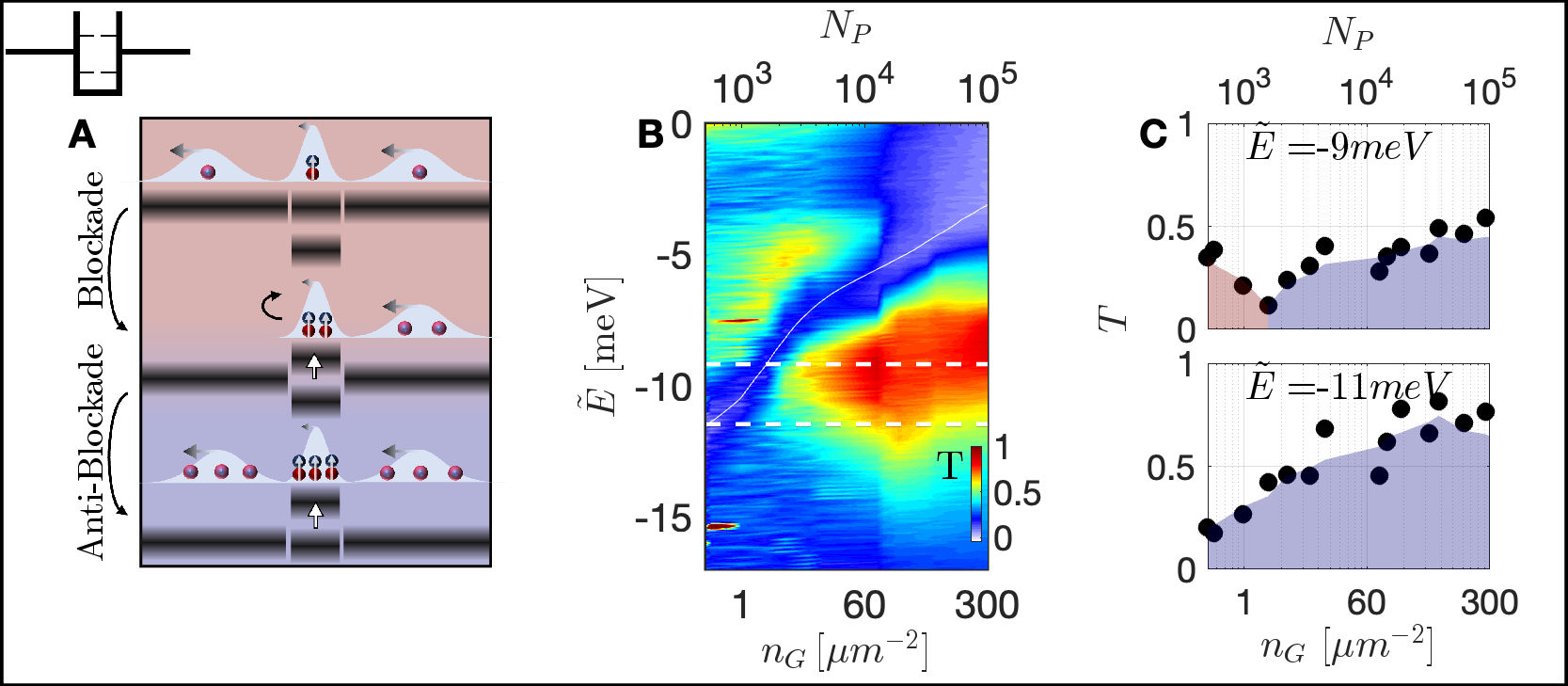}
    \caption{
    \textbf{Blockade and anti-blockade of a small number of dipolaritons}
    \newline
    (\textbf{A}) principle of operation of the polariton blockade and anti-blockade.
    (\textbf{B}) transmission spectrum of the polaritons with a constant gate bias ($\Delta E_X(N_P \rightarrow 0) = - 11.7 meV$) as a function of the density in the gate $n_G$ (bottom axis), and $N_P$ (top axis). 
    (\textbf{C}) Two slices of the transmission spectrum at $\tilde{E} = -9,-12 meV$ respectively as a function of $N_P$ ($n_G$), showing the blockade (pink) and anti-blockade (blue) regimes. The slices are marked in \textbf{B} by white dashed lines.
    } % caption
    \label{fig:Fig4}
\end{figure*}
% \twocolumngrid

\section{Blockade and anti-blockade of dipolaritons}

Fig. \ref{fig:Fig4} demonstrates how our electrically controlled dipolariton transistor can be utilized to display both a polariton blockade and an anti-blockade behavior. 
Fig. \ref{fig:Fig4}A presents schematics of a blockade/anti-blockade: first, the gate is biased such that the MP in the gate overlaps with the LP in the channel and transmits polaritons. Then, $N_P$ is increased, and the slow dipolaritons interact, shifting the states in the gate, such that the LP-MP gap overlaps with the LP branch in the channel, fully blocking transmission. This is the mechanism identified with dipolariton blockade. A further increase of $N_P$ results in an even larger interaction-induced blue-shift of the gate, until full transmission is regained via an LP-LP alignment. This part corresponds to an anti-blockade process, similar to the facilitation shown for Rydberg atoms \cite{Marcuzzi2017FacilitationDisorder}.
In fig. \ref{fig:Fig4}B we plot the transmission spectrum as function of $n_G$ (bottom x-axis, see Eq.(\ref{eq:gate_dens})) and $N_P$ (top x-axis), measured with a constant $F_G$ (corresponding to $\Delta E = -12 meV$ at low $n_G$). Here, we observe a clear blue-shift of the LP-MP gap (white line guides the eye) as $n_G$ increases. 
The blockade/anti-blockade mechanism is presented by two fixed energy cross-sections of this map (marked by the white dashed lines), plotted in \ref{fig:Fig4}C:
In the top panel, the MP in the gate overlap with the LP in the channel at low $N_P$, resulting in a high transmission. When  $N_P$ is increased to only $\sim 1000$, the $T$ drops to a minimum, demonstrating a polariton blockade behavior.
By further increasing $N_P$, $T$ rises again, demonstrating an anti-blockade behavior.
The bottom panel presents a full anti-blockade behavior by again increasing $N_P$ to only $\sim 1000$. 
%*************************%
%%          Table 1.         %%
\begin{table}[htbp]
\begin{center}
\begin{tabular}{ c c c c }
\hline \hline
 & B &AB& \\

\hline
$\tilde{E}$   & $-9$  & $-11$ & meV \\
% $\Delta E_{dd} $     & $2$ & $3$ & meV                   \\
$\Delta n _{B / AB}$ & 2.8   &  5.7 & $\mu m ^{-2}$ \\
$\Delta N _{P}$ &  1000  &  1600  &     \\
% $\bar{\eta}           $ & 10 &  25  &       \\
% $\tau_p\times w$      & $2$  & $2$ & $\mu m \cdot ps$      \\
$\tau_p\times w$      & $1100$  & $1100$ & $\mu m \cdot ps$      \\
\hline
\hline
\end{tabular}
\caption{
\textbf{Towards a 2-dipolariton blockade}
The table presents the values extracted from the two transitions described in figure \ref{fig:Fig4}C, namely the blockade (B) and anti-blockade (AB).
% $\Delta E_{dd}$ is measured by the energy shift of the Rabi gap between the markers, measured from Fig. \ref{fig:Fig4}B.
$\Delta N_P$ and $\Delta n_{B/AB}$ are the corresponding Polaritons number/density difference for blockade/anti-blockade
% $\bar{\eta}$ is the extracted energy averaged density compression ratio defined by $\bar{\eta}  =n_G/n_C$.
$\tau_p \times w$ are the corresponding values of the  pulse duration times the channel width in the current experiment.  
} % caption
\label{tbl1}
\end{center}
\end{table}

Importantly, based on this demonstration, we estimate that both polariton blockade and anti-blockade at the quantum level of only 2-polaritons should be feasible in such a system. 
This can be seen by setting $N_P = 2$ and choosing a well defined energy $E_P$ for an injected polariton pulse in Eq. \ref{eq:gate_dens}. Such substitution simply gives:
$n^G_B = 2/[\tau_P\cdot w \cdot v_{ph} \cdot \eta (E_P, F_G)]$, where $\eta= v_g^G(E_p,F_G)/v_{ph}$. The pulse length $\tau_P$ is limited from below by the polariton spectral width $\gamma_P$ (determined mostly by the excitonic linewidth), $\tau_P>h/\gamma_P$, where in our case $\gamma_{P} \simeq 0.3 meV$ (see Fig. S1). This limits the polariton pulse duration to be $\geq 2.2 ps$. Taking $n^G_B=\Delta n_B$ from table \ref{tbl1}, and substituting the values for $\tau_P$, $v_{ph}$ yields a condition on the maximal width of the optical waveguide mode for a true 2-polariton blockade:

\begin{gather}
w\times\eta\leq 4\times 10^{-3} \mu m.
\end{gather}

For example, using the extracted value $\eta=2.5\times 10^{-3}$ for $\Tilde{E}=-9$ meV gives $w\leq 1.5 \mu m$, a value that should be easily achieved by standard lithography techniques and can safely support guided modes.

While it seems that increasing $\gamma_P$ allows for shorter pulses and thus higher $n_G$, we note that it also smears the transmission response to increasing $n_B^G$ and increases the polariton losses, thus it expected to yield no improvement.

\section{Large dipolar-induced polariton interactions: experiment and theory}

The density-dependent blue shift of the dipolaritons is seen nicely by the shift of the LP-MP gap, marked by a white line in Fig. \ref{fig:Fig4}B.
A linear fit to the shift of the gap at low densities: $\Delta E_{dd}=g_{dd}(F_G)\cdot n_G$ 
yields a distinctively large $g_{dd}(\Delta E_X=-11.7meV) = 0.7\pm 0.2 meV\cdot \mu m^2$, a value very similar to our previous results \cite{Rosenberg2018a}, confirming the significant enhancement of dipolariton interactions over non-polar polaritons or fixed-dipole polaritons \cite{Cristofolini2012,Togan2018EnhancedPolaritons}.

%Taking into consideration the unique electrostatic structure of the studied sample, wherein the dipolar-exciton is confined within a single wide QW, our model enables a strong dipolar-induced screening mechanism
To understand and model this large density-dependent polariton blue-shift, we suggest a simple dielectric screening model. The dielectric screening induced by the polarized exciton part of the polariton can be presented by an effective density-dependent dielectric constant $\epsilon_{dd}(n_{G})$. 
The dipolar screening reduces the electric field in the QW, as given by (see full derivation in Appendix \ref{app:Sceening model}):

\begin{equation}
    E_G=F_G/\epsilon_{dd}(n_{G})=F_G/(1+\chi_{dd}),
\end{equation}
where,
\begin{equation}
    \chi_{dd} = \frac{(N_d/V_d)\cdot d_X}{\epsilon_0 F_G}=\frac{N_d}{V_d}\cdot\frac{\alpha}{\epsilon_0}= \frac{n_G}{L_d} \frac{ \alpha}{\epsilon_0} .
\end{equation}
Here $N_d$ and $V_d$ are the number and volume of the induced excitonic dipoles, $d_X(E_G)=\alpha E_G$, where $\alpha$ is the effective polarizability of an exciton. $L_d(E_G)\approx d_X(E_G)/e$ is the effective dipole layer thickness in the QW quantization direction.
%%          Figure 5     %%

Generally, the energy of free dipoles with a magnitude $d_X=eL_d$ and a  density $n_{G}$ that are introduced into a dielectric medium under a fixed bias $V_G$ is (See Appendix \ref{app:dip inter} for full derivation):
\begin{equation}\label{eq:dipole_int}
    \Delta E_X(F_G,n_G)=-d_X F_G+\frac{d_X n_{G}  e}{\epsilon_0\epsilon},
\end{equation}
where the first term comes from the dipole-field interaction and the second from the uncorrelated dipole-dipole interaction (known as the plate capacitor formula).
Importantly, unlike the case of fixed dipoles, for induced dipoles
$d_X=\alpha E_G=\alpha F_G/\epsilon_{dd}(n_G) = \alpha \frac{F_G}{1+\alpha n_G/\epsilon_0 L_d}$, similar to model in Ref.\cite{Rosenberg2018a} Eq. S18 but derived in a different manner. This leads to the following nonlinear density dependent blue shift:
\begin{equation}
\begin{split}
   \Delta E_{dd}(n_G) & = \Delta E_X(F_G,n_G)-\Delta E_X(F_G,0) \\
   &= \alpha F_G^2\left[1 - \left(\frac{1}{( 1+\frac{\alpha}{\epsilon_0} n_G/L_d)} \right)\right] \\
   & +d \frac{\alpha F_G n_G e}{\epsilon \epsilon_0( 1+\frac{\alpha}{\epsilon_0} n_G/L_d)}.
    \label{eq:Stark_density}
    \end{split}
\end{equation}
Taking the result up to the first order in $n_G$ we get:
\begin{equation}
    \Delta E_{dd}(n_G)\simeq \left(\frac{\alpha^2F_G^2}{L_d\epsilon_0 } + \frac{\alpha F_G e}{\epsilon \epsilon_0}\right)n_G,
    \label{}
\end{equation}
and we can identify:
\begin{equation}
g_{dd}\equiv\frac{\Delta E_{dd}(n_G)}{n_G}\simeq
\frac{e d(F_G)}{\epsilon_0}(1+1/\epsilon).
\end{equation}
As can be seen, for induced dipoles, the largest contribution for $g_{dd}$ comes from the density dependent reduction of $d_X$, a term which is absent for fixed dipoles.

\begin{figure}[!t]
    \centering
    \includegraphics[trim= 0 0 0 0,clip,width = 8.3 cm]{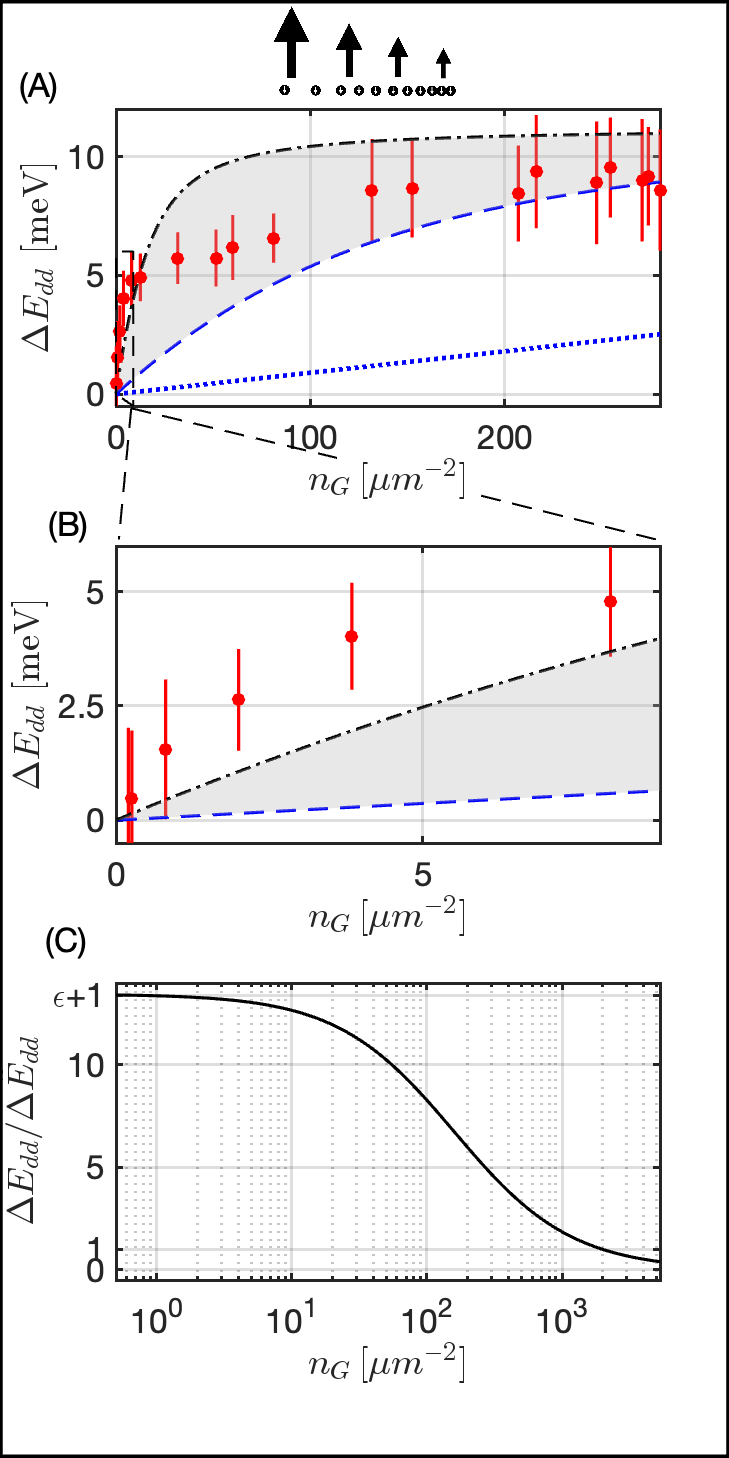}
    \vspace{0ex}
    \caption{
    \textbf{Field-induced dipolariton interactions.}
    \newline
    \textbf{(A)} Measured blue-shift of the LP-MP gap at the gate (red symbols).
    The dashed blue line is the prediction of the polarization screening model, Eq. \ref{eq:Stark_density} (no fitting parameters). The dotted-dashed black line is the prediction for $\chi_{dd}\rightarrow N_{QW} \chi_{dd}$, where $N_{QW} = 8$. For comparison, the bare contact interaction of non-polar excitons is shown by the dotted blue line. The errors here, are the uncertainty in the middle of the LP-MP gap, the bars show the full width at half the depth.
    \textbf{(B)} Zoomed view of the low-density regime of (A).
    \textbf{(C)} The density-dependent interaction-energy blue-shift ratio of the field-induced dipoles to that of fixed dipoles.
    } % caption
    \label{fig:Fig5}
\end{figure}

The only parameter to evaluate in this model is $\alpha$, this can be done by a numerical solution of $\Delta E_X(F_G)$. To do that we first solve the Schrodinger equation of the wide QW under an applied electric field to find the electron and hole energy shifts and wave functions under an applied field \cite{Mazuz-Harpaz2017RadiativeWells,Rosenberg2016}. Then, we use a variational method to calculate the changes in binding energy of the exciton under the same field \cite{Miller1984Band-EdgeEffect,Bastard1988a}. Then $\Delta E_X(F_G,0)$ is calculated as the sum of the single particle shifts and the change of the exciton binding energy without any fitting parameters.
Finally, we set $\alpha=\Delta E_X(F_G,0)/F_G^2$.
A very good agreement was found between the calculated $\Delta E_X(F_G,0)$ and the experimental results, as shown in Fig. \ref{fig:Fig1}B, yielding:
\begin{equation}
 \alpha=1.3 \frac{meV}{\mu m^2/V^2} \text{ , }
\end{equation}
and therefore we have:
\begin{equation}
  g_{dd}= 29\mu eV\mu m^{2}\cdot(\frac{F_G}{F_G=1V/\mu m}),
\end{equation}
which for $F_G=2.9 V/\mu m$ of Fig. \ref{fig:Fig4}, yields $g_{dd}=84 \mu eV \mu m^2$.
\\
The prediction of the model is plotted in Fig. \ref{fig:Fig5}A together with the extracted $\Delta E$ of the LP-MP gap energy from Fig. \ref{fig:Fig4}B. For comparison, the exchange term, $g_0 n_G$, is also plotted. The model prediction nicely follows the experimental data and captures the more than an-order-of-magnitude dipolar interaction enhancement over non-polar polaritons, \textit{with no fitting parameters}. We note that at low densities, the model prediction is lower than the experimental points (see Fig. \ref{fig:Fig5}B). One contribution to the discrepancy is because our model does not take into account the added screening of the induced polarization field of excitons in all adjacent QWs, which is expected to penetrate all across the sample, as the sample thickness is much smaller than the lateral size of the gate. This means that the effective screening field of all the active QWs should be almost additive, $\chi_{dd}\rightarrow N_{QW}\chi_{dd}$, and thus $g_{dd}\rightarrow N_{QW}\cdot g_{dd}$.  In our structure, 8 QWs are essentially in strong coupling with the WG-mode. Setting an effective QW number $N_{QW}\simeq 8$, yields $g_{dd}\simeq 0.67  meV \mu m^2$, within the uncertainty range of the measurements at very low densities.

The effective interaction constant, $g_{dd}$, of field-induced dipoles is different from that of fixed dipoles, $\tilde{g}_{dd}$, where the latter only contains the screened dipole-dipole interaction term in Eq. \ref{eq:dipole_int}, thus yields $\tilde{g}_{dd}=ed_X/\epsilon_0\epsilon$ \cite{Laikhtman2009ExcitonShift,Laikhtman2009CorrelationsInteraction}, so
\begin{equation}
g_{dd}/\tilde{g}_{dd}=\epsilon+1.
\end{equation}
For GaAs QWs, this ratio is $\epsilon_{GaAs}+1=13.4$. Fig. \ref{fig:Fig5}(C) shows the ratio of the field-induced dipole interaction energy of Eq. \ref{eq:Stark_density}, $\Delta E_{dd}$ to that of fixed dipoles $\Tilde{\Delta E_{dd}}$, which is significantly larger than 1 over a wide range of densities. 
This result may explain the mysterious, an-order-of-magnitude discrepancy between the dipolar interaction constant measured for WG-dipolaritons with gated wide single GaAs QWs \cite{Rosenberg2018a,Datta2022HighlyMoS2} and those with asymmetric double GaAs QWs \cite{Cristofolini2012,Togan2018EnhancedPolaritons}. It pinpoints the advantage of using field-induced dipoles for large dipolar nonlinearities, in particular for materials with large dielectric constants.
Importantly, this applies also for polaritons based on interlayer excitons in TMD heterostructures. It suggests that polaritons based on quadrupolar excitons that were predicted \cite{Slobodkin2020QuantumHeterostructures} and recently discovered \cite{Ajit_quad,Heinz_quad,Xie2023BrightHeterotrilayer,Lian2023QuadrupolarSuperlattice,Bai2023EvidenceHeterotrilayer} may be excellent for nonlinear devices due to their field-induced dipole properties.
 \\
Finally, we exclude contributions to the interaction-induced blue shift which are not polaritonic in nature. Since the polaritons are much faster than bare reservoir excitons and free carriers (by order of the mass ratio $(\sim10^4)$), and since the readout takes place at about $100 \mu m$ away from the excitation spot, such a pulsed excitation scheme spatially and temporally separates the polaritons from excitons and other free carriers within each pulse. To prevent pulse-to-pulse memory effects the repetition rate of the laser was selected to be slower than any long living charges in the system (SM Sec. \ref{sec:supp:rep_rate}).
Intra-pulse exciton accumulation under the gate, where the first polaritons arriving at the gate may disassociate or scatter into excitons that later interact with the rest of the polaritons, can also be excluded: the number of charges is limited by the polaritons failing to transmit through the gate, resulting in low densities over the effective gate area ($\sim 120 \mu m^2$), which cannot explain the large observed blue-shifts at high transmission.
We therefore conclude that the observed interactions are very likely polaritonic in nature.

\section{Conclusions}
Strongly interacting gated dipolaritons in wide QWs are a promising platform for the first observation of the long-sought 2-polariton blockade, a crucial building-block for deterministic quantum photonic integrated circuitry.
This demonstration of the fundamental building blocks for electrically controlled photonic logic circuitry in a scalable, planar, on-chip photonic platform approaching the true quantum limit, will open up vast opportunities to realize various types of complex photonic processing, either classical or quantum, using dressed dipolar photons.

\section*{Acknowledgments}
We thank Eldad Betthelheim and Yoad Ordan for helpful discussions. R.R. and D.L. acknowledge the support from the Israeli Science Foundation Grants 836/17 and 1087/22, and from the NSF-BSF Grant 2019737.
J.H., N.L and H.D. acknowledge the support of the National Science Foundation under grant DMR 2004287, the Army Research Office under grant W911NF-17-1-0312, the Air Force Office of Scientific Research under grant FA2386-21-1-4066, and the Gordon and Betty Moore Foundation under grant GBMF10694.
This research is funded in part by the Gordon and Betty Moore Foundation’s EPiQS Initiative, Grant GBMF9615 to L. N. Pfeiffer, and by the National Science Foundation MRSEC grant DMR 2011750 to Princeton University.

%%% appendix...
\renewcommand{\thefigure}{A\arabic{figure}}
\setcounter{figure}{0}% Reset counters
\appendix
\section{Induced dipole screening model }
\label{app:Sceening model}
From an electrostatic point of view, the sample  (see SM for exact details of the sample structure) can be treated as a series of capacitors, each with capacitance $c_i$, hence the total capacitance is given by:
\begin{equation}
C_{tot} = \left(\sum_i \frac{1}{c_i} \right)^{-1} = \left(\sum_i \frac{L_i}{A\epsilon_i} \right)^{-1} = A\frac{\prod_i\epsilon_i}{\sum_jL_j\prod_{i\neq j}\epsilon_i}
\end{equation}
where $L_i$ is the thickness  of each dielectric layer, $\epsilon_i $ is the dielectric constant of each layer and $A$ is its area.
The total capacitance is $Q = C V_0$, where $Q$ is the charge and $V_0$ the applied bias, and the displacement field between the contacts is $D = Q/A$, the displacement field $D=\epsilon E$ is continuous across the layer boundaries. Using these relations and the above equation, the electric field in the QW is given by:
\begin{gather}
E_{QW} = D/\epsilon_{QW} = Q/(A\epsilon_{QW}) = CV_0/(A\epsilon_{QW})
\\
E_{QW} =\frac{V_0}{\epsilon_{QW}} \frac{\prod_i\epsilon_i}{\sum_jL_j\prod_{i\neq j}\epsilon_i} 
\label{eq:supp:Field}
\end{gather}
Now, when excitons are excited in the QW having a background dielectric constant $\epsilon_{QW}$, the electric field polarizes the excitons, resulting in an additional dipolar screening of the external field $\epsilon_{dd}(n_X)$, with a macroscopic polarization $P_X = \epsilon_0\chi_{dd}E_{QW} $. Since $ P_X= n_X d_X = n_X \alpha E_{QW}$ we find:

\begin{equation}
\epsilon_{dd}(n_X)=1+\chi_{dd}(n_X)= 1+\frac{n_X\alpha}{\epsilon_0},
\end{equation}
 where $n_X$ is an effective 3-dimensional density of excitons. This additional screening can be accounted for by replacing $\epsilon_{QW}$ by $\epsilon_{QW}\cdot\epsilon_{dd}$
\begin{gather}
E_{QW} =\frac{V_0}{\epsilon_{QW}\epsilon_{dd}} \frac{\epsilon_{dd}\prod_i\epsilon_i}{\sum_{j\neq QW}L_j\epsilon_{dd}\prod_{i\neq j}\epsilon_i+L_{QW}\prod_{i\neq QW}\epsilon_i} 
\\
E_{QW} =\frac{V_0}{L_{QW}} \frac{1}{\sum_{j\neq QW}\frac{L_j}{L_{QW}}\frac{\prod_{i\neq j}\epsilon_i}{\prod_{i\neq QW}\epsilon_i}\epsilon_{dd}+1} 
\\
E_{QW} =\frac{V_0}{L_{QW}} \frac{1}{\sum_{j\neq QW}\frac{L_j}{L_{QW}}\frac{\epsilon_{QW}}{\epsilon_j}\epsilon_{dd}+1} \equiv \frac{V_0}{L_{QW}} \frac{1}{\gamma\epsilon_{dd}+1},
\end{gather}
where $\gamma \equiv \sum_{j\neq QW}\frac{L_j}{L_{QW}}\frac{\epsilon_{QW}}{\epsilon_j} $ is a property of the structure. Generally, $\gamma\sim L_{S}/L_{QW}$, where $L_{S}$ ($L_{QW}$) are the widths of the sample and QW respectively. In our case therefore $\gamma\gg 1$.
For vanishing dipole density $\epsilon_{dd}\rightarrow 1$ and we can define
\begin{equation}
    E_{QW}(n_X=0)\equiv F_0 = \frac{V_0}{L_{QW}} \frac{1}{\gamma+1}\simeq \frac{V_0}{L_{S}},  
    \label{eq:supp:V2F}
\end{equation}
for the sample under study $F_0 = 0.9 \frac{V_0}{L_{S}} = 0.85 V_0 \, V/\mu m$.
Therefore for a finite dipole density we get 
\begin{equation}
     E_{QW}(n_X) = F_0\frac{\gamma+1}{\epsilon_{dd} \gamma +1}\simeq \frac{F_0}{\epsilon_{dd}}=\frac{F_0}{1+\chi_{dd}}
     \label{eq:supp:screenF}
\end{equation}
\begin{figure}[H]
    \centering
    \includegraphics[width=0.5\linewidth]{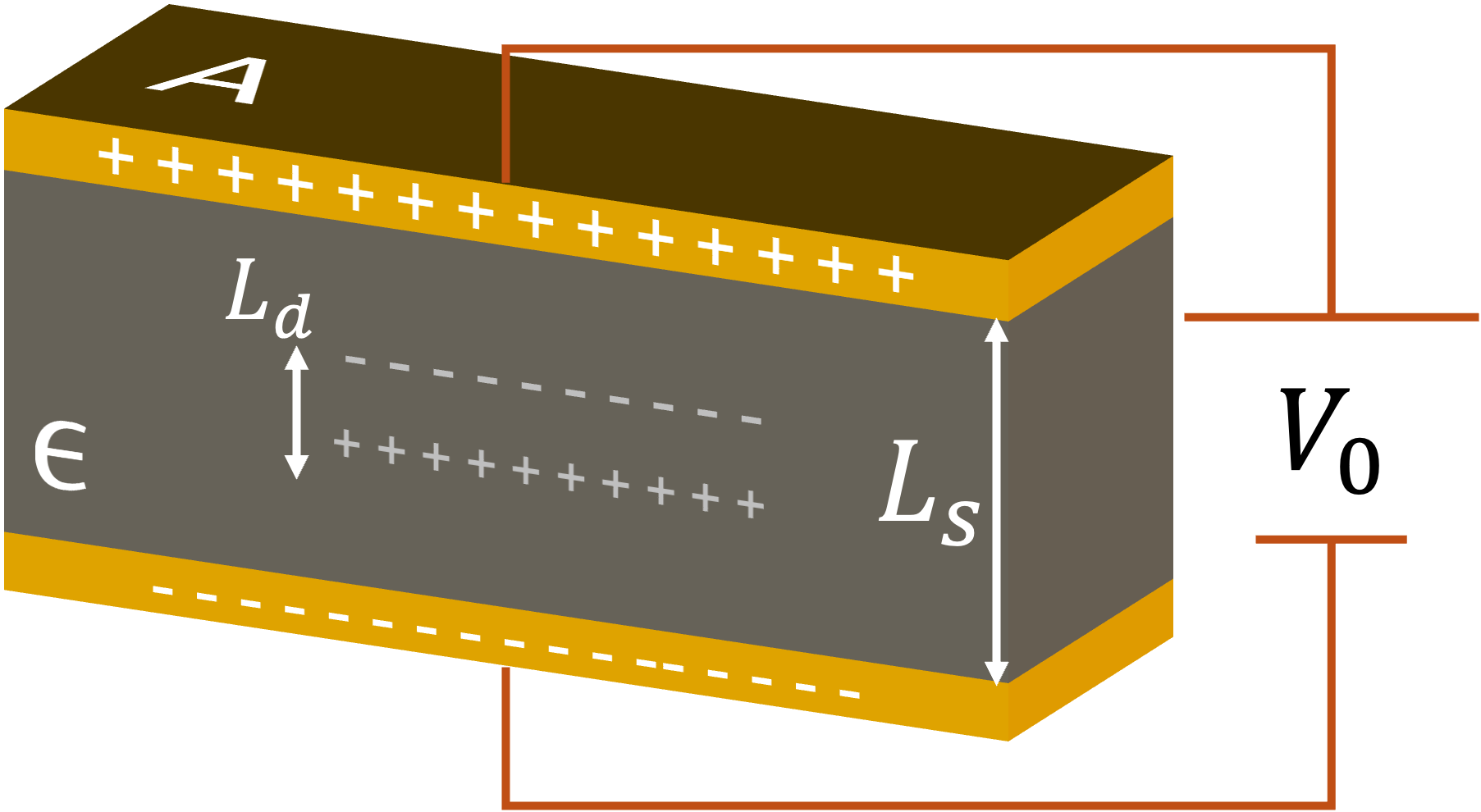}
    \caption{Plate capacitor screening model: illustration.}
    \label{fig:sup:Cpacitor}
\end{figure}

\section{The interaction energy shift of dipolar excitons}
\label{app:dip inter}
The energy of dipoles introduced into an electrically biased dielectric medium $\epsilon$ can be calculated as the energy change in the medium when two oppositely charged layers with a separation $L_d$ and total charge $Q$ are inserted into the biased medium. This change is given by:
\prx{\cite{Jackson1999ClassicalElectrodynamics}}:
\begin{gather}
\Delta U = \frac{1}{2}\int\bold{d}V\left[E_f\cdot D_f\right]-\frac{1}{2}\int\bold{d}V\left[E_i\cdot D_i\right]
\\
\Delta U = \frac{ 1}{2}\int\bold{d}V\left[E_f\cdot D_f - E_i\cdot D_i\right],  
\end{gather}
where $E_i,D_i$ are the fields before without the dipoles and $E_f,D_f$ are with them.
The displacement field $D \equiv \epsilon_0\epsilon E$ outside of the inserted charged layers is unchanged while inside it becomes $D_f = D_i-n_{2D}$ where $n_{2D} = Q/A$ is the charge density over some area $A$.
Now, we can rewrite the system energy as
\begin{gather}
\Delta U = \frac{ 1}{2}\int\bold{d}V\left[(E_i-n_{2D}/\epsilon_0\epsilon)(D_i-n_{2D})- E_i\cdot D_i\right]
\\
\Delta U = \frac{ 1}{2}\int\bold{d}V\left[-2n_{2D} E_i  + n_{2D}^2/\epsilon_0\epsilon \right]
 \end{gather}
integrating the terms gives
\begin{gather}
\Delta U = \frac{ 1}{2}\int\int\bold{d}A\bold{d}z\left[-2Q/A E_i  + Q^2/A^2\epsilon_0\epsilon \right]
\\
\Delta U = \frac{L_d}{2}\left[2Q E_0  -2 Q^2/A\epsilon_0\epsilon \right]
\\
\Delta U = (QL_d)E_i-(QL_d)(Q/A)/\epsilon_0\epsilon
\end{gather}
now, if we define $QL_d = -NeL_d = N d_X$, we can write the energy per particle:
\begin{gather}
\frac{\Delta U}{N} = -d_X E_i+\frac{d_X n_{2D}  e}{\epsilon_0\epsilon}
\label{eq:sup:DipolarEne}
\end{gather}
where $e$ is the elementary charge. The first term is the  stark shift from the dipole-field interaction and the second one is the mean-field, uncorrelated dipole-dipole interaction term (known as the plate capacitor formula), see Ref. \cite{Laikhtman2009CorrelationsInteraction,Laikhtman2009ExcitonShift}
\\
Eq. \ref{eq:sup:DipolarEne} describes the energy for either fixed dipoles ($d_X=d_0$) or for induced dipoles ($d_X=d(E_f)$) where the dipole is a function of the screened electric field $E_f$ (rather than of $E_i$) ,which in turn depends on the dipolar density ($n_{2D}$), as demonstrated in Eq. \ref{eq:supp:screenF}, $d(E(n_X)) = \alpha E_f(n_X) = \alpha E_i/(1+\chi_{dd})$, where $\chi_{dd} = n_X \alpha/\epsilon_0$, $E_i = F_0 = E_{QW} (n_X = 0) $ and $E_f = E_{QW}(n_X)$. As we are interested in a thin layer of dipoles, the density is $n_X = n_{2D}/L_d$, where $L_d = d_X/e$ is the effective thickness of the layer. Explicitly the dipolar energies are given by:
\begin{gather}
    \Tilde{U}_{dd} = -d_0 E_i+\frac{d_0 n_{2D} e}{\epsilon_0\epsilon}
    \\
    U_{dd} = -\frac{\alpha E_i}{1+\chi_{dd}}E_i+\frac{\frac{\alpha E_i }{1+\chi_{dd}} n_{2D}  e}{\epsilon_0\epsilon}
\end{gather}
Here, $\Tilde{U}_{dd}$ is the fixed dipole case, and $U_{dd}$ is for induced dipoles.
Now if we look at the energy shift as the density grows we can write $\Delta E  = U(n)-U(n=0)$
\begin{gather}
    \Delta \Tilde{E}_{dd} = \frac{d_0 n_{2D} e}{\epsilon_0\epsilon}
    \\
    \Delta E_{dd} = \alpha E_i^2 \left[ 1- \frac{1}{1+\chi_{dd}}\right]+\frac{\frac{\alpha E_i }{1+\chi_{dd}} n_{2D}  e}{\epsilon_0\epsilon}
\end{gather}
Taking this results up to first order in $n_{2D}$, and substituting $\chi_{dd}=\frac{\alpha n_{2D}} {L_d \epsilon_0}$ gives:
\begin{gather}\label{eq:induced_dipole_energy}
    \Delta \Tilde{E}_{dd} = \frac{ d_0 e}{\epsilon_0\epsilon}n_{2D}
    \\
    \Delta E_{dd} = \left(\frac{\alpha^2 E_i^2} {L_d \epsilon_0}+\frac{\alpha E_i  e}{\epsilon_0\epsilon}\right)n_{2D}
\end{gather}
We can identify and compare the interaction constants for both cases, $\Tilde{g_{dd}}$ $g_{dd}$, by setting $d_0  = \alpha E_i$ and $ L_d = d_0/e$:
\begin{gather}
    \Tilde{g}_{dd} = \frac{d_0 e}{\epsilon_0\epsilon}
    \\
    g_{dd} = \frac{d_0 e}{\epsilon_0}(1+1/\epsilon).
\end{gather}
Thus, the ratio of the interaction constant for low densities is given by
\begin{gather}
    \frac{g_{dd}}{\tilde{g}_{dd}} = \epsilon+1.
\end{gather}
For dipoles in vacuum ($\epsilon=1$), the ratio of the interaction constants is reduced to $2$ as expected. This is intuitively understood as in this case the two terms in Eq. \ref{eq:induced_dipole_energy}  become equal, and the induced dipole has equal energy contributions from both the dipole-dipole term and the dipole length reduction while for the fixed dipole only the first term contributes.

%*************************%
%%%%%%%%%%%%%%%%%%%%%%%%%%%%%%
\bibliography{references}

\begin{thebibliography}{10}

\bibitem{Wang2019IntegratedTechnologies}
Jianwei Wang, Fabio Sciarrino, Anthony Laing, and Mark~G. Thompson.
\newblock {Integrated photonic quantum technologies}.
\newblock {\em Nature Photonics 2019 14:5}, 14(5):273--284, 10 2019.

\bibitem{Elshaari2020HybridCircuits}
Ali~W. Elshaari, Wolfram Pernice, Kartik Srinivasan, Oliver Benson, and Val Zwiller.
\newblock {Hybrid integrated quantum photonic circuits}.
\newblock {\em Nature Photonics 2020 14:5}, 14(5):285--298, 4 2020.

\bibitem{Bogaerts2020ProgrammableCircuits}
Wim Bogaerts, Daniel P{\'{e}}rez, José Capmany, David~A.B. Miller, Joyce Poon, Dirk Englund, Francesco Morichetti, and Andrea Melloni.
\newblock {Programmable photonic circuits}.
\newblock {\em Nature 2020 586:7828}, 586(7828):207--216, 10 2020.

\bibitem{butovReview2017}
L.~V. Butov.
\newblock {Excitonic devices}.
\newblock {\em Superlattices and Microstructures}, 108:2--26, 8 2017.

\bibitem{ExcitonRevTMD2022}
Alberto Ciarrocchi, Fedele Tagarelli, Ahmet Avsar, and Andras Kis.
\newblock {Excitonic devices with van der Waals heterostructures: valleytronics meets twistronics}.
\newblock {\em Nature Reviews Materials 2022 7:6}, 7(6):449--464, 1 2022.

\bibitem{SAWExcitons2014}
S.~Lazi{\'{c}}, A.~Violante, K.~Cohen, R.~Hey, R.~Rapaport, and P.~V. Santos.
\newblock {Scalable interconnections for remote indirect exciton systems based on acoustic transport}.
\newblock {\em Physical Review B - Condensed Matter and Materials Physics}, 89(8):085313, 2 2014.

\bibitem{ButovTransistor2008}
Alex~A. High, Ekaterina~E. Novitskaya, Leonid~V. Butov, Micah Hanson, and Arthur~C. Gossard.
\newblock {Control of exciton fluxes in an excitonic integrated circuit}.
\newblock {\em Science}, 321(5886):229--231, 7 2008.

\bibitem{Deng2010}
Hui Deng, Hartmut Haug, and Yoshihisa Yamamoto.
\newblock {Exciton-polariton Bose-Einstein condensation}.
\newblock {\em Reviews of Modern Physics}, 82(2):1489--1537, 5 2010.

\bibitem{SanvittoTheRoadDevices}
Daniele Sanvitto and Stéphane K{\'{e}}na-Cohen.
\newblock {The road towards polaritonic devices}.
\newblock {\em Nature Materials}, 15(10):1061--1073, 7 2016.

\bibitem{Chang2018Colloquium:Photons}
D.~E. Chang, J.~S. Douglas, A.~Gonz{\'{a}}lez-Tudela, C.~L. Hung, and H.~J. Kimble.
\newblock {Colloquium: Quantum matter built from nanoscopic lattices of atoms and photons}.
\newblock {\em Reviews of Modern Physics}, 90(3):031002, 8 2018.

\bibitem{Peyronel2012QuantumAtoms}
Thibault Peyronel, Ofer Firstenberg, Qi~Yu Liang, Sebastian Hofferberth, Alexey~V. Gorshkov, Thomas Pohl, Mikhail~D. Lukin, and Vladan Vuleti{\'{c}}.
\newblock {Quantum nonlinear optics with single photons enabled by strongly interacting atoms}.
\newblock {\em Nature 2012 488:7409}, 488(7409):57--60, 7 2012.

\bibitem{Tiarks2018AInteractions}
Daniel Tiarks, Steffen Schmidt-Eberle, Thomas Stolz, Gerhard Rempe, and Stephan D{\"{u}}rr.
\newblock {A photon–photon quantum gate based on Rydberg interactions}.
\newblock {\em Nature Physics 2018 15:2}, 15(2):124--126, 10 2018.

\bibitem{De2023Room-TemperatureCondensates}
Jianbo De, Xuekai Ma, Fan Yin, Jiahuan Ren, Jiannian Yao, Stefan Schumacher, Qing Liao, Hongbing Fu, Guillaume Malpuech, and Dmitry Solnyshkov.
\newblock {Room-Temperature Electrical Field-Enhanced Ultrafast Switch in Organic Microcavity Polariton Condensates}.
\newblock {\em Journal of the American Chemical Society}, 145(3):1557--1563, 1 2023.

\bibitem{Zasedatelev2021Single-photonTemperature}
Anton~V. Zasedatelev, Anton~V. Baranikov, Denis Sannikov, Darius Urbonas, Fabio Scafirimuto, Vladislav~Yu Shishkov, Evgeny~S. Andrianov, Yurii~E. Lozovik, Ullrich Scherf, Thilo St{\"{o}}ferle, Rainer~F. Mahrt, and Pavlos~G. Lagoudakis.
\newblock {Single-photon nonlinearity at room temperature}.
\newblock {\em Nature 2021 597:7877}, 597(7877):493--497, 9 2021.

\bibitem{SpinSwitchAmo2010}
A.~Amo, T.~C.H. Liew, C.~Adrados, R.~Houdr{\'{e}}, E.~Giacobino, A.~V. Kavokin, and A.~Bramati.
\newblock {Exciton-polariton spin switches}.
\newblock {\em Nature Photonics}, 4(6):361--366, 6 2010.

\bibitem{Marsault2015}
Félix Marsault, Hai~Son Nguyen, Dimitrii Tanese, Aristide Lema{\^{i}}tre, Elisabeth Galopin, Isabelle Sagnes, Alberto Amo, and Jacqueline Bloch.
\newblock {Realization of an all optical exciton-polariton router}.
\newblock {\em Applied Physics Letters}, 107(20):201115, 11 2015.

\bibitem{TunnelingDiode2013Bloch}
H.~S. Nguyen, D.~Vishnevsky, C.~Sturm, D.~Tanese, D.~Solnyshkov, E.~Galopin, A.~Lema{\^{i}}tre, I.~Sagnes, A.~Amo, G.~Malpuech, and J.~Bloch.
\newblock {Realization of a Double-Barrier Resonant Tunneling Diode for Cavity Polaritons}.
\newblock {\em Physical Review Letters}, 110(23):236601, 6 2013.

\bibitem{Gao2012Transistor}
T.~Gao, P.~S. Eldridge, T.~C.H. Liew, S.~I. Tsintzos, G.~Stavrinidis, G.~Deligeorgis, Z.~Hatzopoulos, and P.~G. Savvidis.
\newblock {Polariton condensate transistor switch}.
\newblock {\em Physical Review B - Condensed Matter and Materials Physics}, 85(23):235102, 6 2012.

\bibitem{Ballarini2013}
D.~Ballarini, M.~De~Giorgi, E.~Cancellieri, R.~Houdr{\'{e}}, E.~Giacobino, R.~Cingolani, A.~Bramati, G.~Gigli, and D.~Sanvitto.
\newblock {All-optical polariton transistor}.
\newblock {\em Nature Communications}, 4:1778, 4 2013.

\bibitem{Suchomel2017PrototypeSwitch}
H.~Suchomel, S.~Brodbeck, T.~C.H. Liew, M.~Amthor, M.~Klaas, S.~Klembt, M.~Kamp, S.~H{\"{o}}fling, and C.~Schneider.
\newblock {Prototype of a bistable polariton field-effect transistor switch}.
\newblock {\em Scientific Reports 2017 7:1}, 7(1):1--9, 7 2017.

\bibitem{Delteil2019}
Aymeric Delteil, Thomas Fink, Anne Schade, Sven H{\"{o}}fling, Christian Schneider, and Ataç İmamo{\u{g}}lu.
\newblock {Towards polariton blockade of confined exciton–polaritons}.
\newblock {\em Nature Materials}, 18(3):219--222, 3 2019.

\bibitem{Munoz-Matutano2019EmergencePolaritons}
Guillermo Mu{\~{n}}oz-Matutano, Andrew Wood, Mattias Johnsson, Xavier Vidal, Ben~Q. Baragiola, Andreas Reinhard, Aristide Lema{\^{i}}tre, Jacqueline Bloch, Alberto Amo, Gilles Nogues, Benjamin Besga, Maxime Richard, and Thomas Volz.
\newblock {Emergence of quantum correlations from interacting fibre-cavity polaritons}.
\newblock {\em Nature Materials}, 18(3):213--218, 3 2019.

\bibitem{Directinteractions_snoke_2017}
Yongbao Sun, Yoseob Yoon, Mark Steger, Gangqiang Liu, Loren~N. Pfeiffer, Ken West, David~W. Snoke, and Keith~A. Nelson.
\newblock {Direct measurement of polariton-polariton interaction strength}.
\newblock {\em Nature Physics}, 13(9):870--875, 2017.

\bibitem{Snoke2023ReanalysisMicrocavities}
D.~W. Snoke, V.~Hartwell, J.~Beaumariage, S.~Mukherjee, Y.~Yoon, D.~M. Myers, M.~Steger, Z.~Sun, K.~A. Nelson, and L.~N. Pfeiffer.
\newblock {Reanalysis of experimental determinations of polariton-polariton interactions in microcavities}.
\newblock {\em Physical Review B}, 107(16):165302, 4 2023.

\bibitem{Vladimirova2010interactions}
M.~Vladimirova, S.~Cronenberger, D.~Scalbert, K.~V. Kavokin, A.~Miard, A.~Lema{\^{i}}tre, J.~Bloch, D.~Solnyshkov, G.~Malpuech, and A.~V. Kavokin.
\newblock {Polariton-polariton interaction constants in microcavities}.
\newblock {\em Physical Review B - Condensed Matter and Materials Physics}, 82(7):075301, 8 2010.

\bibitem{Rosenberg2016}
Itamar Rosenberg, Yotam Mazuz-Harpaz, Ronen Rapaport, Kenneth West, and Loren Pfeiffer.
\newblock {Electrically controlled mutual interactions of flying waveguide dipolaritons}.
\newblock {\em Physical Review B}, 93(19):195151, 5 2016.

\bibitem{Rosenberg2018a}
Itamar Rosenberg, Dror Liran, Yotam Mazuz-Harpaz, Kenneth West, Loren Pfeiffer, and Ronen Rapaport.
\newblock {Strongly interacting dipolar-polaritons}.
\newblock {\em Science Advances}, 4(10):eaat8880, 10 2018.

\bibitem{Suarez-Forero2021EnhancementInteractions}
D.~G. Su{\'{a}}rez-Forero, F.~Riminucci, V.~Ardizzone, N.~Karpowicz, E.~Maggiolini, G.~Macorini, G.~Lerario, F.~Todisco, M.~De~Giorgi, L.~Dominici, D.~Ballarini, G.~Gigli, A.~S. Lanotte, K.~West, K.~Baldwin, L.~Pfeiffer, and D.~Sanvitto.
\newblock {Enhancement of Parametric Effects in Polariton Waveguides Induced by Dipolar Interactions}.
\newblock {\em Physical Review Letters}, 126(13):137401, 3 2021.

\bibitem{Tsintzos2018ElectricalCondensates}
S.~I. Tsintzos, A.~Tzimis, G.~Stavrinidis, A.~Trifonov, Z.~Hatzopoulos, J.~J. Baumberg, H.~Ohadi, and P.~G. Savvidis.
\newblock {Electrical Tuning of Nonlinearities in Exciton-Polariton Condensates}.
\newblock {\em Physical Review Letters}, 121(3):037401, 7 2018.

\bibitem{Cristofolini2012}
Peter Cristofolini, Gabriel Christmann, Simeon~I. Tsintzos, George Deligeorgis, George Konstantinidis, Zacharias Hatzopoulos, Pavlos~G. Savvidis, and Jeremy~J. Baumberg.
\newblock {Coupling quantum tunneling with cavity photons}.
\newblock {\em Science}, 336(6082):704--707, 5 2012.

\bibitem{Togan2018EnhancedPolaritons}
Emre Togan, Hyang~Tag Lim, Stefan Faelt, Werner Wegscheider, and Atac Imamoglu.
\newblock {Enhanced Interactions between Dipolar Polaritons}.
\newblock {\em Physical Review Letters}, 121(22):227402, 11 2018.

\bibitem{Datta2022HighlyMoS2}
Biswajit Datta, Mandeep Khatoniar, Prathmesh Deshmukh, Félix Thouin, Rezlind Bushati, Simone De~Liberato, Stephane~Kena Cohen, and Vinod~M. Menon.
\newblock {Highly nonlinear dipolar exciton-polaritons in bilayer MoS2}.
\newblock {\em Nature Communications 2022 13:1}, 13(1):1--7, 10 2022.

\bibitem{Walker2017}
P.~M. Walker, L.~Tinkler, B.~Royall, D.~V. Skryabin, I.~Farrer, D.~A. Ritchie, M.~S. Skolnick, and D.~N. Krizhanovskii.
\newblock {Dark Solitons in High Velocity Waveguide Polariton Fluids}.
\newblock {\em Physical Review Letters}, 119(9):097403, 8 2017.

\bibitem{Liran2018a}
Dror Liran, Itamar Rosenberg, Ken West, Loren Pfeiffer, and Ronen Rapaport.
\newblock {Fully Guided Electrically Controlled Exciton Polaritons}.
\newblock {\em ACS Photonics}, 5(11):4249--4252, 2018.

\bibitem{Brimont2020GaN}
C.~Brimont, L.~Doyennette, G.~Kreyder, F.~R{\'{e}}veret, P.~Disseix, F.~M{\'{e}}dard, J.~Leymarie, E.~Cambril, S.~Bouchoule, M.~Gromovyi, B.~Alloing, S.~Rennesson, F.~Semond, J.~Z{\'{u}}{\~{n}}iga-P{\'{e}}rez, and T.~Guillet.
\newblock {Strong Coupling of Exciton-Polaritons in a Bulk Ga N Planar Waveguide: Quantifying the Coupling Strength}.
\newblock {\em Physical Review Applied}, 14(5):054060, 11 2020.

\bibitem{SuarezLaser2020}
D.~G. Su{\'{a}}rez-Forero, F.~Riminucci, F.~Riminucci, V.~Ardizzone, M.~De Giorgi, L.~Dominici, F.~Todisco, G.~Lerario, L.~N. Pfeiffer, G.~Gigli, D.~Ballarini, D.~Sanvitto, and D.~Sanvitto.
\newblock {Electrically controlled waveguide polariton laser}.
\newblock {\em Optica, Vol. 7, Issue 11, pp. 1579-1586}, 7(11):1579--1586, 11 2020.

\bibitem{Nigro2022IntegratedInterferometry}
Davide Nigro, Vincenzo D’Ambrosio, Daniele Sanvitto, and Dario Gerace.
\newblock {Integrated quantum polariton interferometry}.
\newblock {\em Communications Physics 2022 5:1}, 5(1):1--10, 2 2022.

\bibitem{Christensen2022MicroscopicDipolaritons}
Esben~R. Christensen, Arturo Camacho-Guardian, Ovidiu Cotlet, Atac Imamoglu, Michiel Wouters, Georg~M. Bruun, and Iacopo Carusotto.
\newblock {Microscopic theory of cavity-enhanced interactions of dipolaritons}.
\newblock {\em arXiv}, 2212.02597v1, 12 2022.

\bibitem{Ling2023DeeplyNanophotonics}
Haonan Ling, Arnab Manna, Jialiang Shen, Ho-Ting Tung, David Sharp, Johannes Fr{\"{o}}ch, Johannes Fr{\"{o}}ch, Siyuan Dai, Siyuan Dai, Arka Majumdar, Arka Majumdar, Arka Majumdar, and Artur~R. Davoyan.
\newblock {Deeply subwavelength integrated excitonic van der Waals nanophotonics}.
\newblock {\em Optica, Vol. 10, Issue 10, pp. 1345-1352}, 10(10):1345--1352, 10 2023.

\bibitem{Marcuzzi2017FacilitationDisorder}
Matteo Marcuzzi, Jiří Min{\'{a}}{\v{r}}, Daniel Barredo, Sylvain De~L{\'{e}}s{\'{e}}leuc, Henning Labuhn, Thierry Lahaye, Antoine Browaeys, Emanuele Levi, and Igor Lesanovsky.
\newblock {Facilitation Dynamics and Localization Phenomena in Rydberg Lattice Gases with Position Disorder}.
\newblock {\em Physical Review Letters}, 118(6):063606, 2 2017.

\bibitem{Mazuz-Harpaz2017RadiativeWells}
Yotam Mazuz-Harpaz, Kobi Cohen, Boris Laikhtman, Ronen Rapaport, Ken West, and Loren~N. Pfeiffer.
\newblock {Radiative lifetimes of dipolar excitons in double quantum wells}.
\newblock {\em Physical Review B}, 95(15):155302, 4 2017.

\bibitem{Miller1984Band-EdgeEffect}
D.~A.~B. Miller, D.~S. Chemla, T.~C. Damen, A.~C. Gossard, W.~Wiegmann, T.~H. Wood, and C.~A. Burrus.
\newblock {Band-Edge Electroabsorption in Quantum Well Structures: The Quantum-Confined Stark Effect}.
\newblock {\em Physical Review Letters}, 53(22):2173--2176, 11 1984.

\bibitem{Bastard1988a}
G.~Bastard.
\newblock {\em {Wave mechanics applied to semiconductor heterostructures}}.
\newblock Les Editions de Physique, 1988.

\bibitem{Laikhtman2009ExcitonShift}
B.~Laikhtman and R.~Rapaport.
\newblock {Exciton correlations in coupled quantum wells and their luminescence blue shift}.
\newblock {\em Physical Review B - Condensed Matter and Materials Physics}, 80(19):195313, 11 2009.

\bibitem{Laikhtman2009CorrelationsInteraction}
B.~Laikhtman and R.~Rapaport.
\newblock {Correlations in a two-dimensional Bose gas with long-range interaction}.
\newblock {\em EPL (Europhysics Letters)}, 87(2):27010, 7 2009.

\bibitem{Slobodkin2020QuantumHeterostructures}
Yevgeny Slobodkin, Yotam Mazuz-Harpaz, Sivan Refaely-Abramson, Snir Gazit, Hadar Steinberg, and Ronen Rapaport.
\newblock {Quantum Phase Transitions of Trilayer Excitons in Atomically Thin Heterostructures}.
\newblock {\em Physical Review Letters}, 125(25):255301, 12 2020.

\bibitem{Ajit_quad}
Weijie Li, Zach Hadjri, Luka~M. Devenica, Jin Zhang, Song Liu, James Hone, Kenji Watanabe, Takashi Taniguchi, Angel Rubio, and Ajit Srivastava.
\newblock {Quadrupolar–dipolar excitonic transition in a tunnel-coupled van der Waals heterotrilayer}.
\newblock {\em Nature Materials 2023 22:12}, 22(12):1478--1484, 10 2023.

\bibitem{Heinz_quad}
Leo Yu, Kateryna Pistunova, Jenny Hu, Kenji Watanabe, Takashi Taniguchi, and Tony~F. Heinz.
\newblock {Observation of quadrupolar and dipolar excitons in a semiconductor heterotrilayer}.
\newblock {\em Nature Materials}, 22(12):1485--1491, 12 2023.

\bibitem{Xie2023BrightHeterotrilayer}
Yongzhi Xie, Yuchen Gao, Fengyu Chen, Yunkun Wang, Jun Mao, Qinyun Liu, Saisai Chu, Hong Yang, Yu~Ye, Qihuang Gong, Ji~Feng, and Yunan Gao.
\newblock {Bright and Dark Quadrupolar Excitons in the WSe2/MoSe2/WSe2 Heterotrilayer}.
\newblock {\em Physical Review Letters}, 131(18):186901, 11 2023.

\bibitem{Lian2023QuadrupolarSuperlattice}
Zhen Lian, Dongxue Chen, Lei Ma, Yuze Meng, Ying Su, Li~Yan, Xiong Huang, Qiran Wu, Xinyue Chen, Mark Blei, Takashi Taniguchi, Kenji Watanabe, Sefaattin Tongay, Chuanwei Zhang, Yong~Tao Cui, and Su~Fei Shi.
\newblock {Quadrupolar excitons and hybridized interlayer Mott insulator in a trilayer moir{\'{e}} superlattice}.
\newblock {\em Nature Communications 2023 14:1}, 14(1):1--7, 8 2023.

\bibitem{Bai2023EvidenceHeterotrilayer}
Yusong Bai, Yiliu Li, Song Liu, Yinjie Guo, Jordan Pack, Jue Wang, Cory~R. Dean, James Hone, and Xiaoyang Zhu.
\newblock {Evidence for Exciton Crystals in a 2D Semiconductor Heterotrilayer}.
\newblock {\em Nano Letters}, 23(24):11621--11629, 12 2023.

\bibitem{Jackson1999ClassicalElectrodynamics}
John~David Jackson.
\newblock {\em {Classical electrodynamics}}.
\newblock wiley, 3rd edition, 1999.

\bibitem{1995ConfinedPhotons}
Elias Burstein and Claude Weisbuch.
\newblock {\em {Confined Electrons and Photons}}, volume 340 of {\em NATO ASI Series}.
\newblock Springer US, Boston, MA, 1995.

\bibitem{Mahmut2023VoigtCentral}
Ruzi Mahmut.
\newblock {voigt line shape fit - File Exchange - MATLAB Central}, 2023.

\bibitem{Https://www.filmetrics.com/refractive-index-databaseMeasurement}
{https://www.filmetrics.com/refractive-index-database – Table of Refractive Index Values for Thin Film Thickness Measurement}.

\bibitem{Iotti1997CrossoverWells}
Rita~Claudia Iotti and Lucio~Claudio Andreani.
\newblock {Crossover from strong to weak confinement for excitons in shallow or narrow quantum wells}.
\newblock {\em Physical Review B}, 56(7):3922, 8 1997.

\end{thebibliography}
\bibliographystyle{unsrt}
\nocite{1995ConfinedPhotons}
\nocite{Mahmut2023VoigtCentral}
\nocite{Https://www.filmetrics.com/refractive-index-databaseMeasurement}
\nocite{Iotti1997CrossoverWells}
% \nocite{Laikhtman2009ExcitonShift}
% \nocite{Tassone1999Exciton-excitonPolaritons}

\end{document}

%*************************%
%% Authors and Affiliations
\title{Supplementary Material for Electrically controlled photonic circuits of field-induced dipolaritons with huge nonlinearities}
%  
%*************************%
%% 
\author{Dror Liran}
\email{dror.liran@mail.huji.ac.il}
\author{Ronen Rapaport}
\affiliation{Racah Institute of Physics, The Hebrew University of Jerusalem,  Jerusalem 9190401, Israel}
\author{Jiaqi Hu}
\author{Nathanial Lydick}
\author{Hui Deng}
\affiliation{University of Michigan, Ann Arbor, MI 48109, USA}
\author{Loren Pfeiffer}
\affiliation{Department of Electrical Engineering, Princeton University, Princeton, NJ, 08544 USA}

\maketitle
%\onecolumngrid
%\begin{multicols}{2}
%******%
The supplementary material is arranged as follows: in Sec. I we describe the details of the models used to calculate the dispersion of WG-polaritons under bias, and the derivation of the transmission coefficients through electrical discontinuities. In Section II we show the effect of the repetition rate of the laser on the slow reservoir charge and exciton accumulation, and the method for selecting the repetition rate to avoid the reservoir effects on measurements. In Sec. III we show the details of the density calibration methods, and in Sec. IV we present the screening model as the source of the induced dipolar exciton nonlinearities. Finally we present the sample details in Sec. V

\section{Triple oscillator model and calculated polariton dispersion under voltage and transmission through potential discontinuities} \label{theo_ovlp}
%******%
\subsection{Three coupled oscillators}\label{3osc}
\begin{figure*}
    \centering
\includegraphics[width =\textwidth]{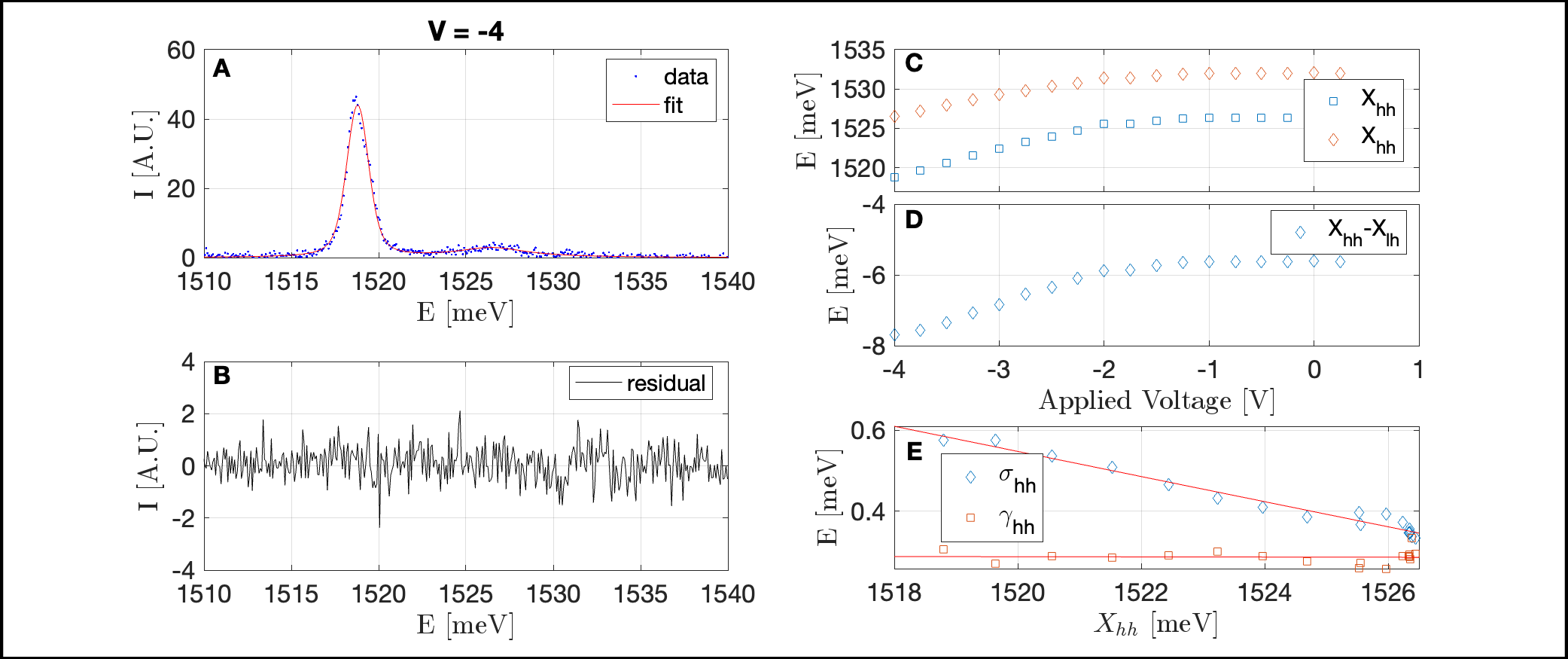}
    \caption{
    \textbf{Measured parameters of the excitons}
    \newline
    The parameters in this figure were extracted by fitting the exciton spectrum to a voigt profile \cite{Mahmut2023VoigtCentral}.
    \textbf{A} Example of fitting the spectrum of the bare excitons with a voigt profile fit.
    \textbf{B} Residue graph for the fit in A.
    \textbf{C} The excitons ($X_{hh},X_{lh}$) energy as function of the applied voltage.
    \textbf{D} The energy difference in between the two excitons as function of the applied voltage.
    \textbf{E} The energy width of the Lorentzian ($\gamma_{hh}$) and Gaussian ($\sigma_{hh}$) parts of the voigt profile.
    } % caption
    \label{fig:sup:voigt}
\end{figure*}

The spectrum of the WG-polariton system can be described using a three coupled oscillator model:
\begin{gather}
    \mathcal{H} = \left(
    \begin{matrix}
    X_{hh} & 0 & \Omega_{h}\\
    0 & X_{lh} & \Omega_{l}\\
    \Omega_{h} & \Omega_{l} & E_{ph}
    \end{matrix}
    \right)
    \\
    E_{ph} = \underbrace{ \hbar c/n}_p\sqrt{k_z^2+\beta^2}
\end{gather}
where both the heavy-hole (with energy $X_{hh}$) and the light-hole (with energy $X_{lh}$) excitons are coupled to the TE WG-photon (with energy $E_{ph}$) with a coupling strength $\Omega_{l/h}$.
The number of free parameters in this model can be reduced using a prior knowledge: the energy separation of the excitons $\Delta_{l-h} = X_{lh}-X_{hh}$ can be measured directly, and it is plotted in fig. \ref{fig:sup:voigt}D. The ratio of the Rabi-coupling of the TE photon with the different excitons corresponds to the square root of their oscillator strength ratio. This is given by $f_h:f_l=3:1$ \cite{1995ConfinedPhotons}.
Thus, reduces the model to 4 free parameters ($ X_{hh} , \Omega , p , k_z $)
\begin{gather}
    \mathcal{H}(\beta) = \left(
    \begin{matrix}
    X_{hh} & 0 & \Omega_{h}\\
    0 & X_{hh}+\Delta_{l-h} & \Omega_{h}/\sqrt{3}\\
    \Omega_{h} & \Omega_{h}/\sqrt{3} & p\sqrt{k_z^2-\beta}
    \end{matrix}
    \right)
\end{gather}
For fitting the measured polariton dispersion, we extract the weighted energy for each wave-vector from the experimental data, and perform best fit of the above model to this reduced extracted dispersion.
%******%

%*************************%
%%          Figure voigt     %%
%************************%

\subsection{Model for polariton transmission through potential discontinuity}
\begin{figure*}
    \centering
    \includegraphics[width = \textwidth]{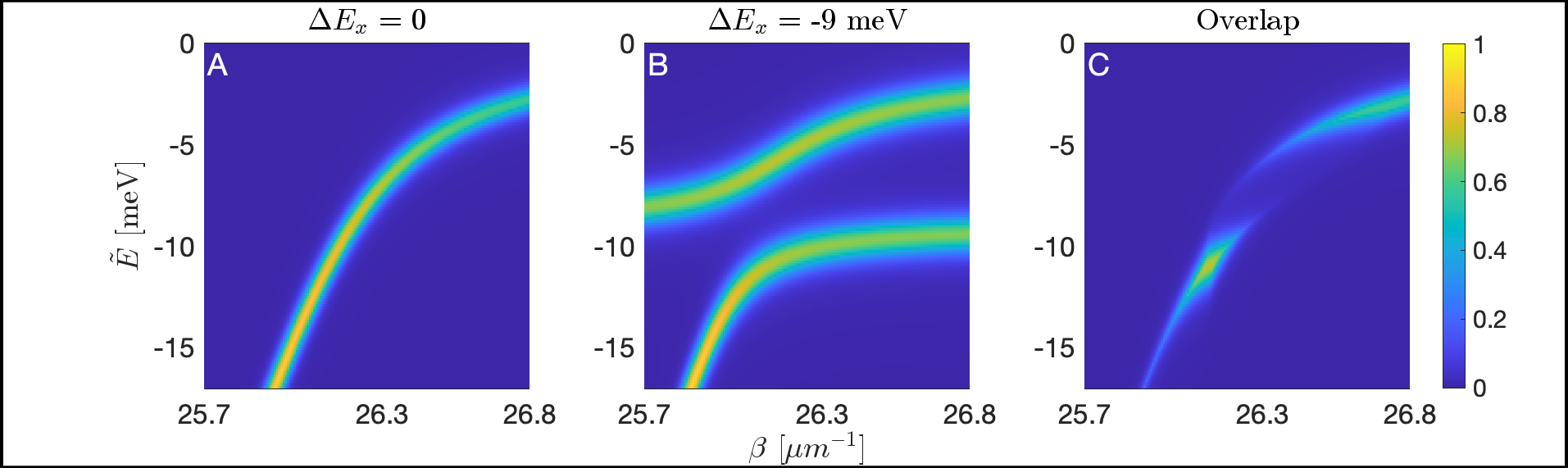}
    \caption{
    \textbf{Generated polariton dispersion}
    \newline
    \textbf{A,B} calculated polariton dispersions at the two sides of a potential discontinuity, with a step of $\Delta E_x = -9 meV$.
    \textbf{C}  the overlap of the two calculated dispersions.
    } % caption
    \label{fig:sup:gen}
\end{figure*}

To model the polariton transmission, we assume that the transmission through the potential discontinuity conserves energy and momentum, thus transmission is only allowed for incoming states which conserve $E(\beta)$ on both sides of the discontinuity. To do this, we take into consideration the energy width for each momentum state. The width of the polaritons can be calculated using the following oscillator model:
\begin{gather}
    \mathcal{H}(\beta) = \left(
    \begin{matrix}
    X_{hh}+i\alpha_{x} & 0 & \Omega_{h}\\
    0 & X_{hh}+\Delta_{l-h}+i\alpha_{x} & \Omega_{h}/\sqrt{3}\\
    \Omega_{h} & \Omega_{h}/\sqrt{3} & p\sqrt{k_z^2-\beta^2}+i\alpha_p
    \end{matrix}
    \right)
\end{gather}
where $\alpha_i$ is the width of the corresponding constituent of the polariton modes, and results from convoluted values of homogeneous (Lorentzian) and inhomogeneous (Gaussian) broadening, which were extracted experimentally.
The photon broadening is described by an extracted voigt profile from the photonic part of the dispersion, with an inhomogeneous (homogeneous) broadening of $\sigma_{ph}  = 1 meV$ ($\gamma_{ph} = 0.05 meV$). The exciton in-homogeneous broadening ($\sigma_{hh}$) varies with the electric field, while the homogeneous broadening ($\gamma_{hh}$) stays constant as is plotted in fig. \ref{fig:sup:voigt}E.
Figure \ref{fig:sup:gen} illustrates an exemplary calculated dispersion of polaritons on the two sides of a discontinuity and their overlap, which then by integrating over $\beta$ yields the transmission $T(\tilde{E})$ due to the potential discontinuity in our model.

%*************************%
%%          Figure generated     %%

\section{Repetition rate} \label{sec:supp:rep_rate}
\begin{figure*}
    \centering
    \includegraphics[width = 0.95\textwidth]{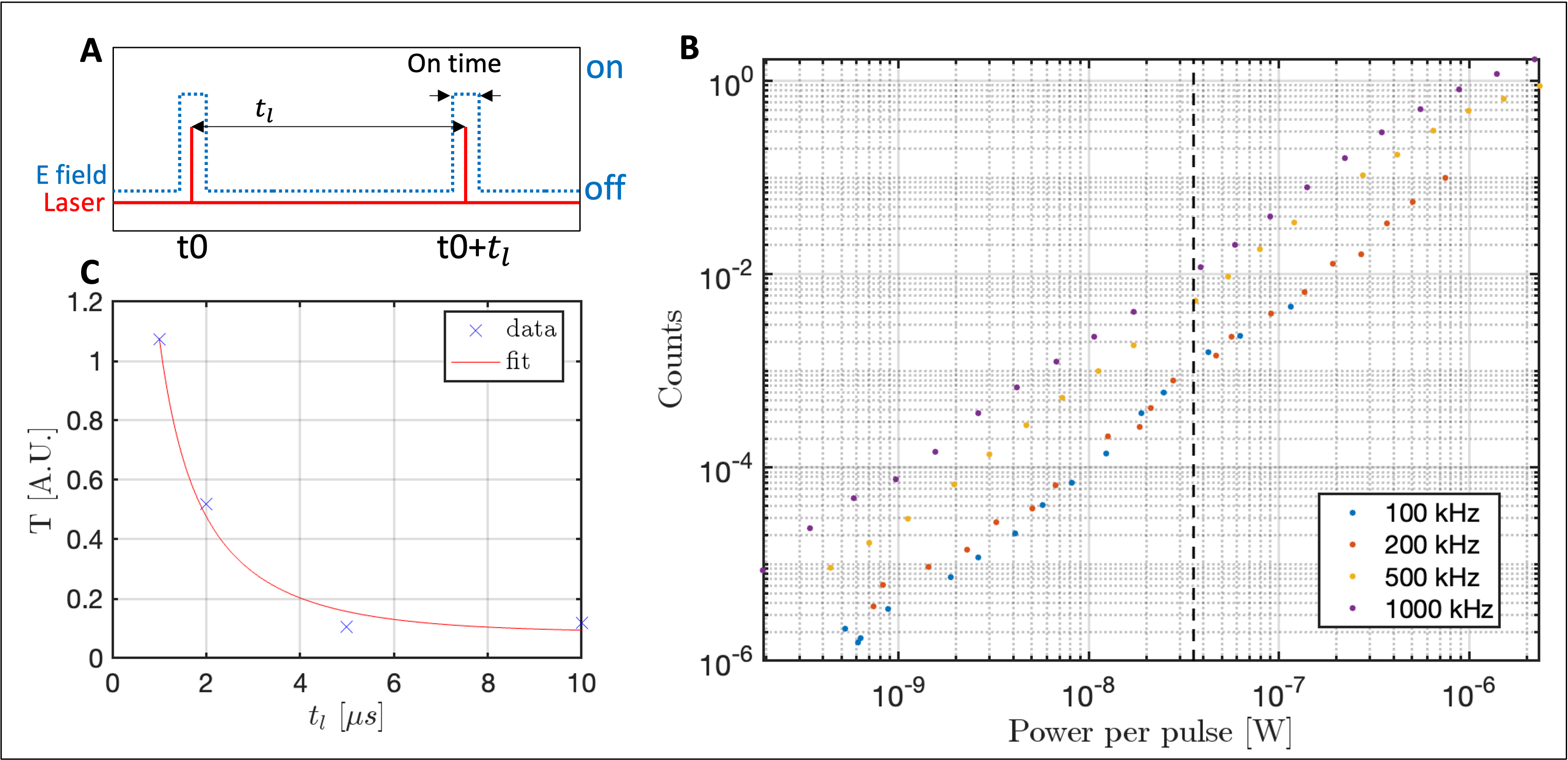}
    \caption{
    \textbf{Laser repetition rate}
    \newline
    \textbf{A} the excitation timing scheme, the electric field turns on $0.25\mu s$ before the laser pulse and turns off $0.25\mu s$ after it. 
    \textbf{B} counts per pulse on the CCD as function of the power per pulse, at repetition rates of 100,200,500,1000 kHz.
    The 100 and 200 kHz measurements overlap.
    \textbf{C} plots a slice of the results from b, plotting counts per-pulse as function of the repetition rate. The data is fitted with eq. \ref{eq:rep}, indicating the long lived carriers live for $\tau_r = 2.4\pm0.2 \mu s$.
    The points are interpolated for power per-pulse for the power marked by dashed line in B. 
    } % caption
    \label{fig:sup:rep}
\end{figure*}

In order to ensure that all of the observed effects are purely polaritonic in nature, and not due to exciton or charge population, we preformed pulsed experiments, as described in the main text. 
Two time scales are relevant here: the fast one, related to the laser pulse length, which starts at the pulse arrival time to the sample and last for few hundreds of pico-seconds until all the polaritons have left the sample. As explained in the main text and in Ref. \cite{Rosenberg2018a} the very different velocities of the polariton with respect to free excitons and free charges and the length of the WG create a built-in space-time separation mechanism within each pulse. The slow one is the longest decay time of any long lived carriers in the system. This is required to be shorter than the pulse to pulse time $t_l$ (the inverse of the laser repetition rate) in order to avoid pulse to pulse memory effects that will mix with the pure polaritonic effects. We measure this slow time scale by preforming transmission experiments through the gated device, similar to the experiment explained in fig. 3, for different repetition rates of the laser, $t_l$, see fig. \ref{fig:sup:rep}A,B.

For repetition rates larger than 200kHz, the total transmission of the polaritons through the gate is seen to depend on the repetition rate, indicating a memory effect from pulse to pulse due to carrier accumulation. For repetition rates equal or smaller than 200kHz, the results are independent on the laser rep rate, indicating that all long living carriers are gone.

To quantify this, the following model was used: the polariton transmission probability through the gate is determined by the induced blue-shift in the gate, a direct result of the total density at the gate at a given time. This implies that the transmission probability goes as:
\begin{gather}
    T(t) \propto \Delta E_G(t) = a N_{P}(t)+b N_X(t)
\end{gather} . $a,b$ include the details of the interaction of each specie (polaritons, exciton and carriers) and their values are not relevant for this estimation.
The number of polaritons is given by
\begin{gather}
    N_{P}(t) = N_1\exp[-t^2/\tau_{p}^2]
\end{gather}
The number of long lived carriers is given by
\begin{gather}
    N_X(t) = N_2\times\sum_{n=-1}^{n=-\infty}\exp(-(t-nt_l)/\tau_r)
    %=N_2\exp(-\gamma_rt)\frac{\exp(-\gamma_rT)}{1-\exp(-\gamma_rT)}
\end{gather}
which means that the transmission probability in time is given by
\begin{gather}
    T(t=0) \propto aN_1+bN_2\frac{1}{\exp(t_l/\tau_r)-1}
    \label{eq:rep}
\end{gather}
in fig. \ref{fig:sup:rep}C we fit the data with Eq. \ref{eq:rep}, which results a charge accumulation life time of $\tau_r = 2.4\pm0.2 \mu s$. Based on these measurements and analysis, we have chosen to do our experiments with a laser repetition rate of $t_l = 200 kHz > 2\tau _r$, so that all memory effects between pulses are negligible.
%******%
%*************************%
%%          Figure rep     %%
%*************************%

\section{Density Calibration} \label{sec:supp:dens_cali}
The polariton density is defined by the number of polaritons for a given area, which in our case is the area of a propagating pulse, $N_P/A_{pulse}$.
We start by counting the number of the injected polaritons, based on the laser power, and the out-coupled polaritons, based on the
measured photons on the
 CCD counts. Then we estimate the pulse area of the propagating pulse.
 
\begin{figure}
    \centering
    \includegraphics[width = 0.5\textwidth]{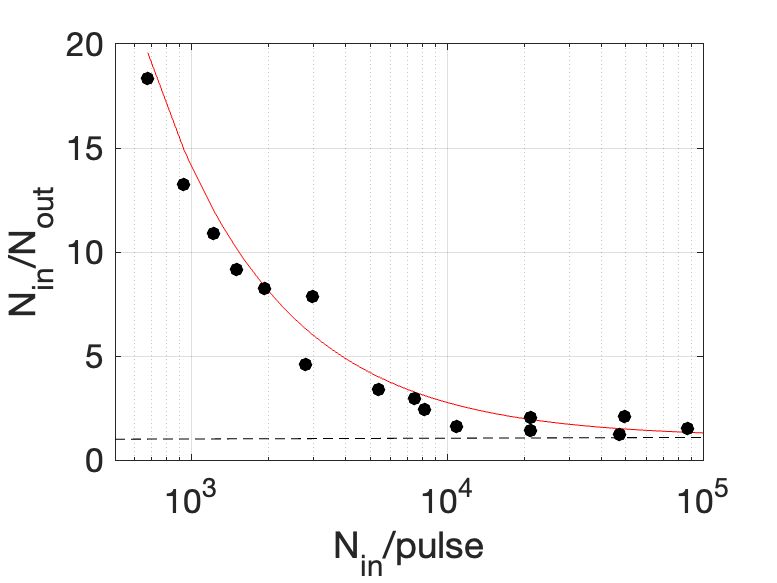}
    \caption{
    \textbf{Number of polaritons}
    The ratio of the to injected photons ($N_{in}$) to emitted photons ($N_{out})$) as function of the injected photons. The red line is a fit, and the dashed black line is the ratio asymptote ($R_{asm}=1.1$).
    } % caption
    \label{fig:sup:inout}
\end{figure}

The number of photons injected and emitted does not exhibit a linear trend in the whole power range of $40 n W-20 \mu W$. While at high powers the input and output counts are linearly proportional to each other, at low powers the output counts are sublinear with the input photons, fig. \ref{fig:sup:inout} plots the ratio. We associate the discrepancy with the low SNR at low powers. Fitting the ratio in fig. \ref{fig:sup:inout} reveals an asymptotic value of $R_{asm} = 1.1$.
Finally, the number of polaritons is calculated as $N_P = N_{in}/R_{asm} $. 

%*************************%
%%          Figure setup     %%
%*************************%

\subsection{Estimating the number of in-coupled polaritons}
An upper bound on the number of injected polaritons can be obtained by the number of photons per pulse, determined by the laser power, and factored by the probability for creating a polariton.
Two main factors are considered in this estimation: the absorption coefficient and the probability of emitting a photon into a supported waveguide mode.

The absorption coefficients is interpreted here as the likelihood of creating an exciton at the injection point. It is calculated with a Transfer matrix formalism, which considers the transmission and reflection resulting from the refractive index mismatch. Additionally, the absorption of the QWs is determined with a Lorentz oscillator model,
the background refractive index is adopted from ref. \cite{Https://www.filmetrics.com/refractive-index-databaseMeasurement} and the oscillator strength was adopted from ref. \cite{Iotti1997CrossoverWells}. This calculation yields a total absorption 
of $\alpha_{abs} = 0.25$ for the laser wavelength ($\lambda  = 774 nm$).

After the excitons are formed in the QWs, they emit in an isotropic angle. Only the photons emitted within the solid angle supported by the waveguide contribute to the polariton population. These angles are $\theta<\bar\theta_c $ and $\phi < \bar\phi_c $. Therefore, the fraction of excitons emmiting into the relevant solid angle are given by eq. \ref{eq:solidA}
\begin{gather}
\frac{1}{4\pi}\int_{-\bar\theta_c^x}^{\bar\theta_c^x} d\phi \int_{\frac{\pi}{2}-\bar\theta_c^{z}}^{\frac{\pi}{2}+\bar\theta_c^{z}} \sin\theta d\theta
\\
= \frac{\bar\theta_c^x\sin(\bar\theta_c^{z})}{\pi} = 3.14\times 10^{-3}
    \label{eq:solidA}
\end{gather}
Using the above values we find the photon to polariton conversion efficiency is $0.785\times 10^{-3}$.

%*************************%
%%          Figure setup     %%
\begin{figure*}
    \centering
    \includegraphics[width = \textwidth]{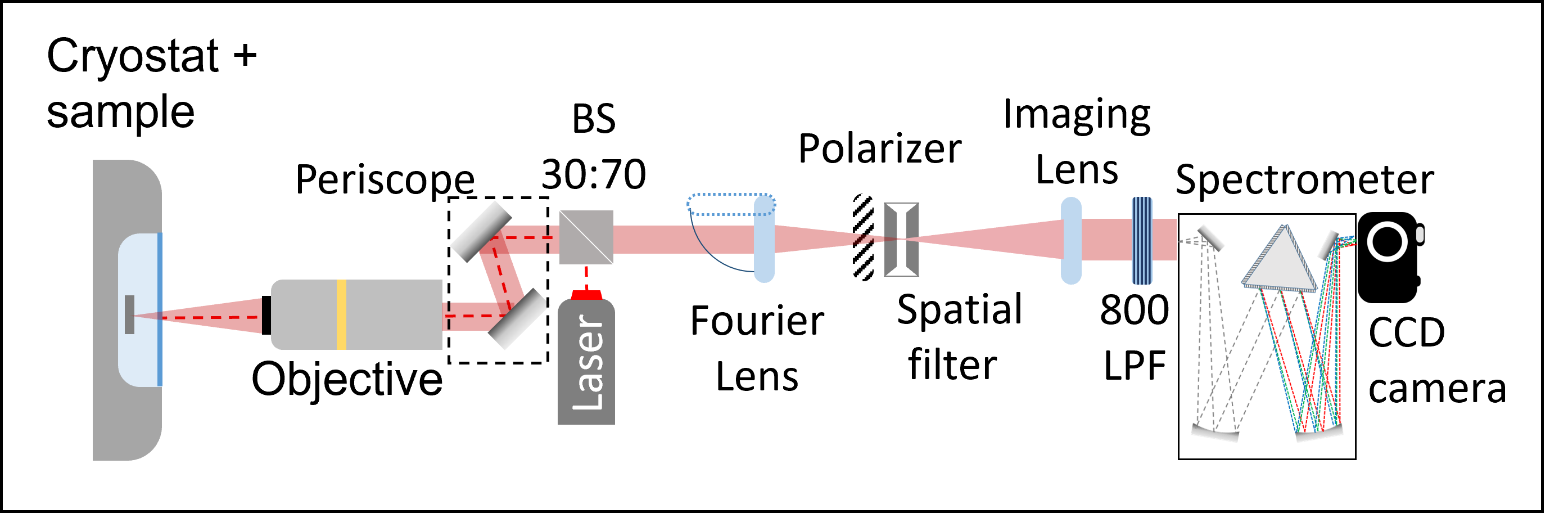}
    \caption{
    \textbf{The optical setup}
    } % caption
    \label{fig:sup:setup}
\end{figure*}
%*************************%

\subsection{Estimating the number of out-coupled polaritons}
The number of out-coupled polaritons per pulse is calculated from the total counts factored by the inverse of total optical loss of the imaging system $\zeta_{system} = \prod_i\zeta_i$, where $\zeta_i$ are the optical loss for each of the elements. The imaging system is illustrated in fig. \ref{fig:sup:setup}. 
When a polariton pulse arrives at the grating coupler it partially couples upwards (into the optical system) and partially downwards (into the substrate), with a power ratio of 1:4.5 respectively, yielding a factor of $\zeta_{grating} = 0.18$. % 1/5.5.
The light is then collected by the objective lens with $NA = 0.2$ which adds an additional factor of $\zeta_{NA} = 0.42$, calculated by eq.\ref{eq:sup:zeta_obj}
\begin{gather}
\zeta_{NA}^{-1} =\frac{\int_0^\infty f(\theta)\textbf{d}\theta}{\int_0^{\theta_{NA}} f(\theta)\textbf{d}\theta} = 2.4
    \label{eq:sup:zeta_obj}
\end{gather}
$f(\theta)$ is the functional distribution of the emission in Fourier space, it is fully defined in eq. \ref{eq:sup:f_theta}. In addition, the loss from the objective corresponds to $\zeta_{objective} = 0.64$
The light then passes through a periscope with two silver mirrors which adds a factor of $\zeta_{periscope} = 0.95$, and through a $30:70 \, (R:T)$  beam splitter which adds a factor of $\zeta_{BS} = 0.63$. Next, it is imaged by two lenses, adding a factor of $\zeta_{lenses} = 0.93$, last before the spectrometer, we positioned an $800 nm$ long pass filter (LPF) to block the Laser light, this adds a factor of $\zeta_{LPF} = 0.97$. 
When the light arrives at the spectrometer it is cut by the slit, yielding a factor of $\zeta_{slit-real} = 0.43, \zeta_{slit-Fourier} = 0.08$ for real or Fourier space configuration respectively, an elaborate discussion follows. All of these factors are summarized in table. \ref{tbl:sup:loss}
The efficiency of the spectrometer and the CCD was measured as a whole: we shined a Laser into the spectrometer and read the counts per incident power, the total conversion efficiency is $142\pm 5 $ photons per count, which translates into $\zeta_{spec+CCD} \simeq 7\times10^{-3}$.
\begin{table}[!ht]
    \centering
        \begin{tabular}{|c|c|}
        \hline
           Element  & factor \\
        \hline
           Grating Coupler  &  0.18\\
           NA           &   0.42 \\
           Objective  &  0.64\\
           Periscope &  0.95 \\
           Beam splitter  & 0.63 \\
           Lenses  & 0.93 \\
           $800 nm$ LPF  & 0.97 \\
           Slit (real space) & 0.43\\
           Slit (Fourier space) & 0.08\\
           Spectrometer and CCD & 7e-3 \\
        \hline
        \end{tabular} \\
    \caption{Summary of the losses of the optical elements}
    \label{tbl:sup:loss}
\end{table}

%*************************%
%%          Figure Real space slit calibration     %%
\begin{figure*}
    \centering
    \includegraphics[width = \textwidth]{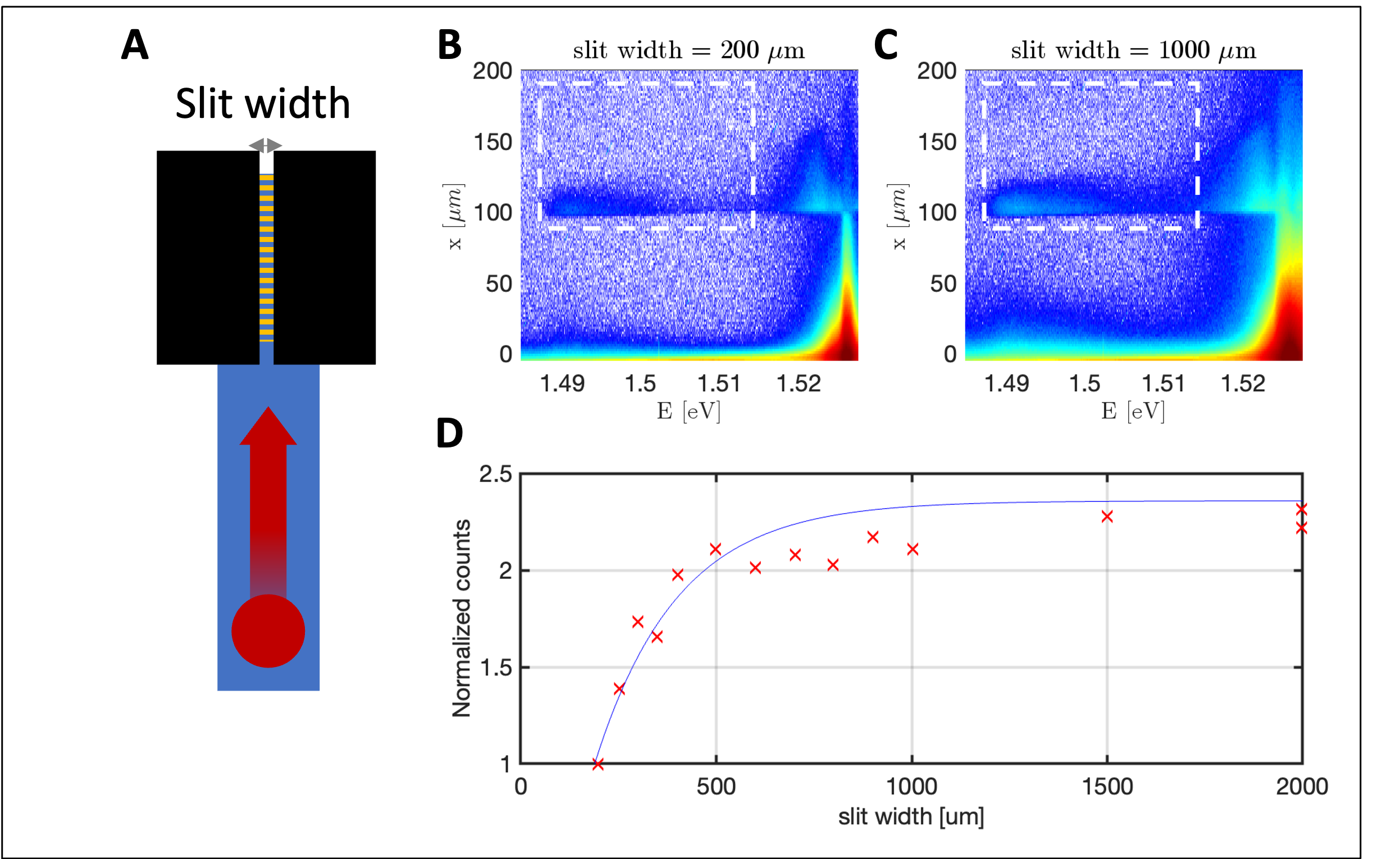}
    \caption{
    \textbf{Real space calibration of the signal fraction on the slit}
    \newline
    \textbf{A} schematic of the experiment, the readout grating is imaged on the slit.
    \textbf{B,C} real space spectrum with slit widths of $200,1000 \, \mu m$.
    \textbf{D} integrated counts as function of the slit opening, the counts are collected from the white dashed rectangle plotted in b and c.
    The data is fitted with eq. \ref{eq:sup:Rslit}, and result a saturation value of 2.35.
    } % caption
    \label{fig:sup:Rslit}
\end{figure*}
%*************************%
\subsubsection{Real space - slit}
To calibrate for the photons that are cut out by the slit in real-space imaging configuration, we measured the dependence of the counts on the opening of the slit, see fig. \ref{fig:sup:Rslit}. The polaritons are counted at energies far from the energy of the exciton, as marked in fig. \ref{fig:sup:Rslit}B,C.
The counts start to saturate at slit widths of about $500 \mu m$, and the exact saturation value is extracted by fitting with eq.  \ref{eq:sup:Rslit}.
\begin{equation}
f(x) = a(1-exp(x/b))+c
\label{eq:sup:Rslit}    
\end{equation}
the fit is plotted in \ref{fig:sup:Rslit}d and yields a saturation value of $2.35$. This gives a factor of $\zeta_{slit-real} = 0.43$.

%*************************%
%%          Figure K space slit calibration     %%
\begin{figure*}
    \centering
    \includegraphics[width = \textwidth]{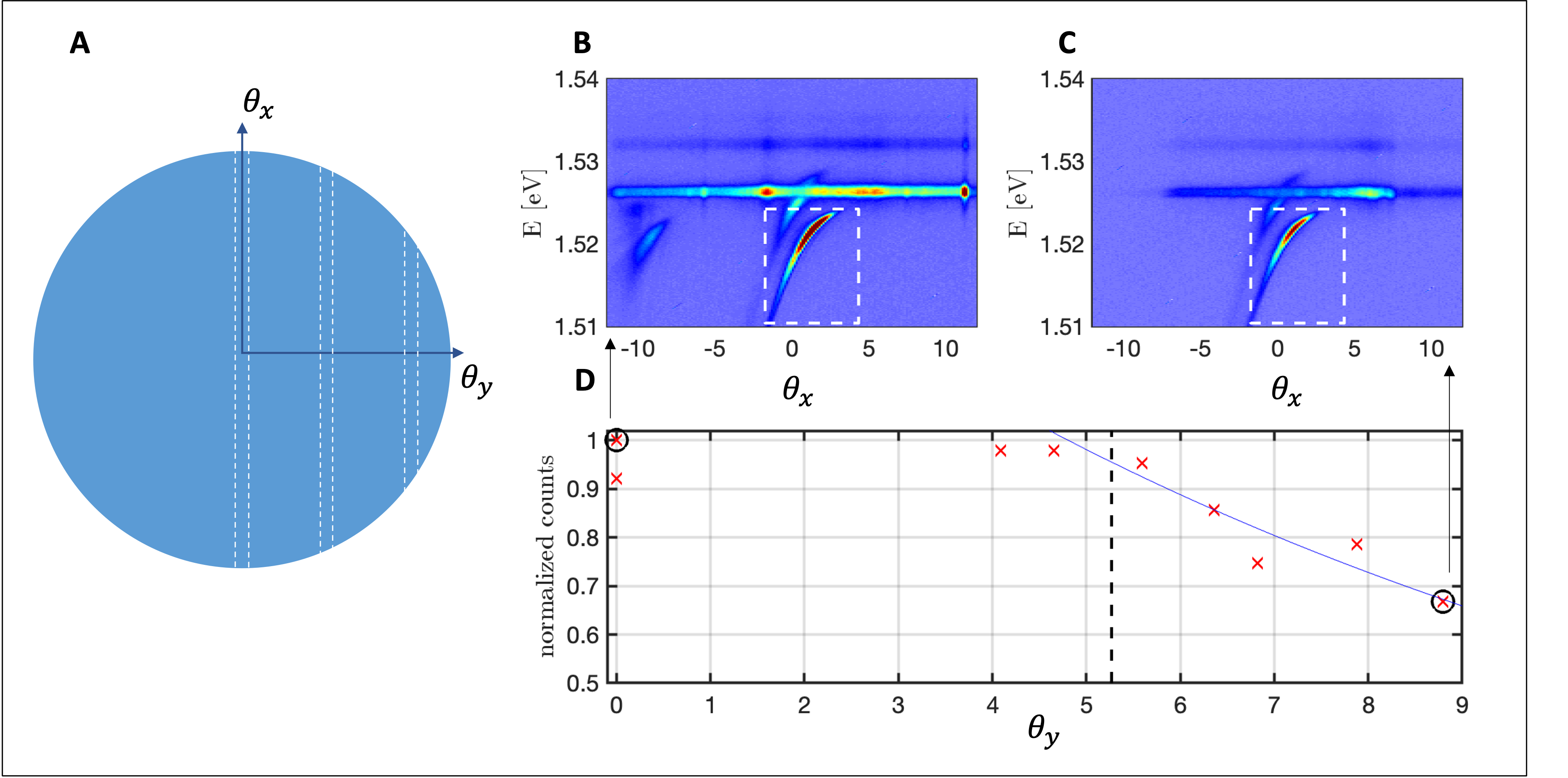}
    \caption{
    \textbf{Fourier space calibration of the signal cut on the slit}
    \newline
    \textbf{A} illustration of the fourier space, the slit is along the $\theta_x$ axis, marked by two parallel white lines. Each image in the set takes a different cut of the fourier space, perpendicular to the slit.
    \textbf{B,C} Fourier space spectrum at $\theta_y \simeq 0,8.8^\circ$.
    \textbf{D} integrated counts as function of the $\theta_y$ displacement, the counts are collected from the white dashed rectangle plotted in b and c.
    } % caption
    \label{fig:sup:Kslit}
\end{figure*}
%*************************%
\subsubsection{Fourier space - slit} \label{sup:FourierSlit}
To calibrate for the photons that are cut on the slit in Fourier space configuration, we have imaged different cuts of the Fourier space, each cut at a different $\theta_y$ value while measuring along $\theta_x$ , see illustration in fig. \ref{fig:sup:Kslit}A.
To estimate the number of polaritons emitted to each angle, we integrated the signal, fig. \ref{fig:sup:Kslit}B,C marks the area of the integration.
When $\theta_y> 5.25^\circ$ we observe a drop in the counts, this drop corresponds to the angle of total internal critical in-plane, results from the lateral confinement.
For angles greater than $\theta_y = 5.25$ the data was fitted with exponential decay, resulting in a characteristic of decay of $\theta_0  = 10.5^\circ$.
Finally, the functional distribution of the emission in Fourier space is given by eq. \ref{eq:sup:f_theta} 
\begin{equation}
    f(\theta) =\begin{cases}
    1 & 0<\theta_y<5.25 \\
    \exp((\theta-5.25)/\theta_0) & 5.25<\theta \\
    \end{cases}
    \label{eq:sup:f_theta}
\end{equation}
integrating over eq. \ref{eq:sup:f_theta} from 0 to $\theta_{NA}$ equal to 12.2, the integration from $\theta_{NA}$ on is already included in $\zeta_{NA}$. Therefore, the relevant factor is $12.2/0.95 = 13$, $0.95^\circ$ is the angle span for the slit opening of $200 \mu m$. Finally $\zeta_{slit-fourier} = 0.08$. 
%*************************%
%%          Figure pulse area     %%
\begin{figure*}
    \centering
    \includegraphics[width = \textwidth]{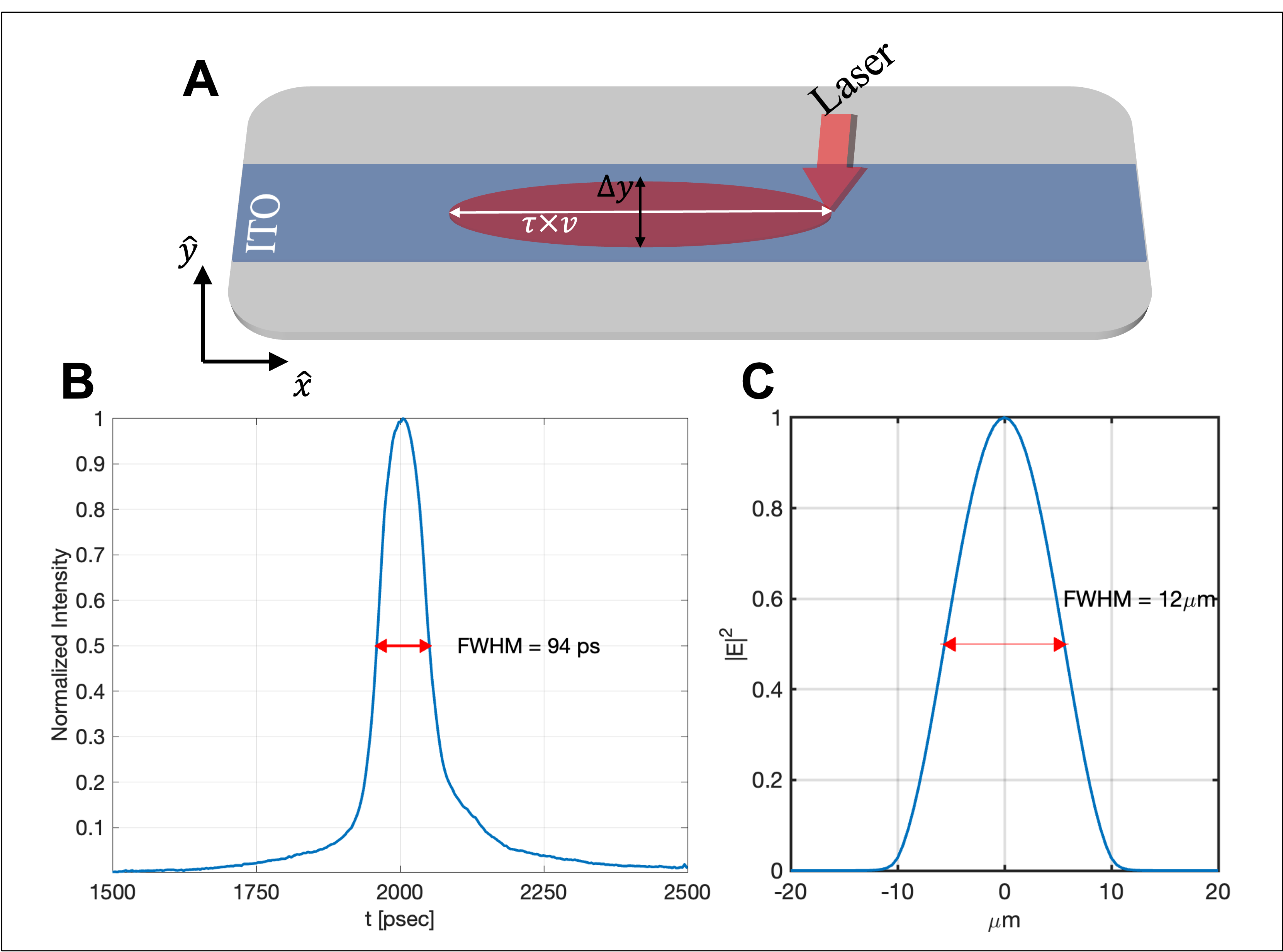}
    \caption{
    \textbf{Pulse area estimation}
    \newline
    \textbf{A} illustration of the pulse area
    \textbf{B} measurement of the pulse duration with a streak camera, the FWHM is 94 pico-seconds.
    \textbf{C} Electric field distribution of the fundamental optical mode in the waveguide, in the lateral direction. The mode was calculated with Finite-Difference-Element solver of Maxwell equations, using Lumerical commercial software.
    } % caption
    \label{fig:sup:pulse_area}
\end{figure*}
%*************************%
\subsection{Estimation of the pulse area} \label{supp:density}
The polariton pulse area is defined by the effective width of the optical mode ($w$) multiplied by the longitudinal spreading of the pulse.
The effective width of the optical pulse is taken as the FWHM of the electromagnetic field distrbution, see fig. \ref{fig:sup:pulse_area}C.
The longitudinal dimension is defined by the pulse duration multiplied by the group velocity $\Delta x \equiv \tau_p \times v_g(E)$.  See fig. \ref{fig:sup:pulse_area}B
The total density is therefore
\begin{equation}
    n = \int dE \frac {N_{pol}(E)}{w\cdot \tau_p v_g^C(E)}    
    \label{eq:sup:density_ch}
\end{equation}
where $w = 12 \mu m; \, \tau_p =94 ps$ ( see fig. \ref{fig:sup:pulse_area}) , and $v_g(E)$is given by the derivative of the energy dispersion ($E_{LP/MP}(\beta)$) with respect to the momentum ($\beta$):
\begin{equation}
    v_g(E) = \frac{\textbf{d}E_{LP/MP}(\beta)}{\hbar \textbf{d}\beta}
    \label{eq:sup:group_vel}
\end{equation}
specifically, in this work the group velocity is derived from the derivative of the fit of a measured dispersion from a channel biased to have the same exciton energy as in the gate, i.e. the same effective field.
\subsubsection{Polariton density under the gate}
As the gate of the transistor device is independently biased, its group velocity dispersion is different than the channel, as is discussed in the main text and presented in fig. 3E. This results in squeezing of the polaritons, by the ratio of the group velocities  $v_{g}^C(E)/v_{g}^G(E)$. The density under the gate is
\begin{equation}
    n_g = \int dE \frac {N_{pol}(E)}{w\cdot \tau_p v_c(E)}\frac{v_g^C(E)}{v_g^G(E)}
    \\
    = \int dE \frac {N_{pol}(E)}{w\cdot \tau_p v_g^G(E)}
\end{equation}
We note that this is a lower limit estimate for the pulse area, which means a upper limit for the density.

\section{Sample details}
\begin{figure}
    \centering
    \includegraphics[width = 0.3\textwidth]{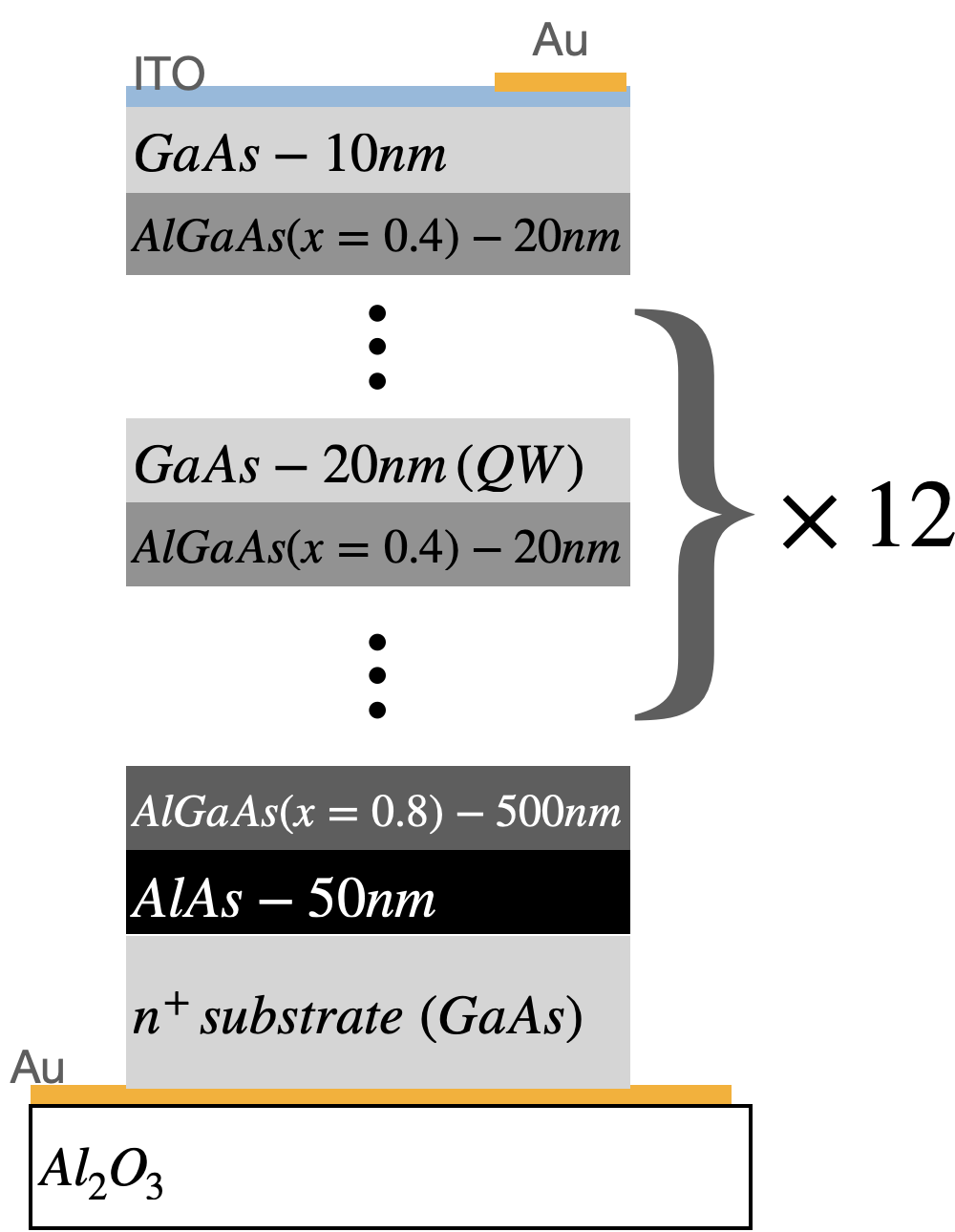}
    \caption{
    \textbf{The layer structure of the sample}
    } % caption
    \label{fig:sample}
\end{figure}

The sample core is made up of twelve $GaAs$ quantum wells (QWs), each of which is $20 nm$ thick. The QWs are separated by $20 nm$ barriers of $Al_xGa_{1-x}As \, (x=0.4)$, giving the entire core a thickness of $500 nm$. The core is grown on top of a $500 nm$ thick clad made of $Al_xGa_{1-x}As \, (x=0.8)$, which sits atop a $50 nm$ stop etch layer of $AlAs$. All of this is grown using molecular beam epitaxy (MBE) on top of an $n^+$ doped $GaAs$ substrate. Additionally, a $10 nm$ $GaAs$ cap layer is grown on top of the sample to prevent oxidation. The full structure is shown in fig. \ref{fig:sample}.

After the MBE growth, we coat the sample with a $120 nm$ of PMMA (950A2), which is patterned by a $500 pA/100kV$ electron beam using the ELS-G100 by ELIONIX. The PMMA is exposed to a dose of $1500 \mu C / cm^2$ in order to write the grating couplers. We then develop the PMMA at $-5^\circ C$ for 1 minute. Next, we deposit a $10 nm$ Ti and $50 nm$ Au layer and remove the unwanted areas using a lift off process. We then deposit a $50 nm$ thick ITO layer and pattern it using laser lithography with AZ1505 photo-resist. The patterned ITO is wet etched using $H_2O:HCl$ (1:1 ratio). Additionally, $100 nm$ thick Au contacts are evaporated onto the sample, for the wire bonding. Finally, we attach the sample to an Al2O3 chip whose top is coated with gold.

%*************************%
%%          Figure sample     %%
%*************************%

\newpage
\bibliography{references}
\bibliographystyle{Science}